\documentclass[aps,pre,superscriptaddress,showpacs,showkeys,nofootinbib]{revtex4-1}

\usepackage{amssymb,amsmath,amsfonts}
\usepackage{graphicx}
\usepackage{color}
\usepackage{url}

\usepackage[latin1]{inputenc}
\usepackage{tikz}
\usetikzlibrary{shapes,arrows}

\newcommand{\calA}{{\mathcal A}}
\newcommand{\calB}{{\mathcal B}}
\newcommand{\calD}{{\mathcal D}}
\newcommand{\calG}{{\mathcal G}}

\newcommand{\wl}{{\overline{w}}}
\newcommand{\dt}{{\Delta t}}
\newcommand{\csch}{{\mbox{csch}}}

\begin{document}
\title{The Kinetics of Wealth and the\\ Origin of the Pareto Law
}
\thanks{\copyright 2012, all rights reserved}
\author{Bruce M. Boghosian}
\affiliation{College of Science and Engineering, American University of Armenia, 40 Baghramian Avenue, Yerevan 0019, Republic of Armenia}
\affiliation{Department of Mathematics, Tufts University, Medford, Massachusetts 02155, USA}
\date{\today}
\begin{abstract}
An important class of economic models involve agents whose wealth changes due to transactions with other agents. Several authors have pointed out an analogy with kinetic theory, which describes molecules whose momentum and energy changes due to interactions with other molecules. We pursue this analogy and derive a Boltzmann equation for the time evolution of the wealth distribution of a population of agents for the so-called Yard-Sale Model of wealth exchange. We examine the solutions to this equation by a combination of analytical and numerical methods, and investigate its long-time limit. We study an important limit of this equation for small transaction sizes, and derive a partial integrodifferential equation governing the evolution of the wealth distribution in a closed economy. We then describe how this model may be extended to include features such as inflation, production and taxation.  In particular, we show that the model with taxation is capable of explaining the basic features of the Pareto law, namely a lower cutoff to the wealth density at small values of wealth, and approximate power-law behavior at large values of wealth.
\end{abstract}
\maketitle


\section{Introduction}
\label{sec:intro}
\subsection{Historical motivation}

In a perfect world the field of economics would not be divided into macroeconomics and microeconomics. The former would be derivable from the latter.  Our current understanding of economics is reminiscent of the situation in statistical physics prior to the 1870s, when the well established field of thermodynamics had to be reconciled with the new atomic theory. The work of Boltzmann, Gibbs, Maxwell and others eventually achieved this reconciliation for dilute gases, demonstrating that the ``macro'' theory of thermodynamics is derivable from the ``micro'' atomic theory.  Economists are still seeking this kind of unification in their field of study.

The knowledge that economics is still incomplete has led some economists to take extreme positions.  There is a school of thought that maintains that no paper on macroeconomics is worth publishing if it is not demonstrably grounded on ``microfoundations''~\cite{bib:Krugman}.  At the same time, over the course of the past twenty-five years, there has been widespread recognition that the very foundations of neoclassical economics -- and microeconomics in particular -- are deeply flawed.  For example, economic agents do not always have perfect information, buyers and sellers do not always behave rationally or even in their own best interests, prices are not always set by an auction process, and it is sometimes not possible to purchase insurance to cover every eventuality.  This has led to a backlash against the ``microfoundations'' proponents that is best summarized in the words of the economist Paul Krugman~\cite{bib:Krugman}, ``$\ldots$the notion that macro is rotten but micro is in good shape is, well, only half right.''

As one might expect, the current situation provides some impetus for transplanting ideas from physics to economics, in the hope that the success of the former subject can be replicated in the latter.  This was the goal of a now-famous meeting at the Santa Fe Institute in 1987 that brought together Nobel laureates in both subjects for this purpose.  The field of ``econophysics'' was arguably born at this meeting, and much progress has been made in the years since.  An outline of the history of the field is described in Beinhocker's book on the subject~\cite{bib:Beinhocker}, and its recent developments have been broken down by country in a very informative recent review journal~\cite{bib:ScienceAndCulture}.

An observation made by numerous authors (see, for example, Ref.~\cite{bib:Yakovenko:2009jt}) is that a useful analogy can be made with the early work of Boltzmann. When molecules collide, they exchange momentum and energy; when economic agents transact, they exchange wealth. If Boltzmann's equation describes the former process, then something similar to a Boltzmann equation ought to describe the latter. This paper pursues this analogy.

There are, of course, essential differences between molecules and economic agents.  For example, in Boltzmann's theory of the former, energy is shared amongst the molecules in a Maxwell-Boltzmann distribution.  There are many hypotheses for the distribution of wealth in societies, and, while some of them involve the Maxwell-Boltzmann distribution in various limits, none are really that simple.

One of the first attempts to quantify the distribution of wealth in a society was made by Vilfredo Pareto in the early twentieth century~\cite{bib:Pareto}.  He studied the distribution of land ownership in Italy by plotting the fraction of people with wealth greater than $x$ versus $x$.  It is clear from the definition of this curve that it is a non-increasing function of $x$.  If we suppose that wealth is distributed according to a probability density function (PDF) $P(w)$, so that $\int_a^b dw\; P(w)$ is the total population with wealth $w\in[a,b]$, then the function that Pareto plotted was
\begin{equation}
A(w) := \frac{1}{N}\int_w^{\infty} dw'\;P(w'),
\label{eq:ParetoA0}
\end{equation}
where $N:=\int_0^\infty dw\;P(w)$ is the total population.  Differentiating both sides of this relation yields
\begin{equation}
P(w) = -N\frac{dA(w)}{dw},
\label{eq:ftoc}
\end{equation}
so the PDF may be easily recovered from Pareto's function.

Pareto found empirically that $A(w)$ was well approximated by
\begin{equation}
A_p(w) \approx
\left\{
\begin{array}{ll}
1 & \mbox{if $w<w_{\min}$}\\
\left(\frac{w_{\min}}{w}\right)^{\alpha} & \mbox{otherwise,}
\end{array}
\right.
\label{eq:ParetoA}
\end{equation}
where $w_{\min}$ is a lower bound on wealth, and the exponent $\alpha$ is called the {\it Pareto index}.  If the total wealth $W:=\int_0^\infty dw\;P(w)w$ of the population is to be finite, it must be that $\alpha>1$.  Using Eq.~(\ref{eq:ftoc}), we find the corresponding Pareto PDF,
\begin{equation}
P_p(w) \approx
\left\{
\begin{array}{ll}
0 & \mbox{if $w<w_{\min}$}\\
\frac{\alpha N}{w_{\min}} \left(\frac{w_{\min}}{w}\right)^{\alpha+1} & \mbox{otherwise.}
\end{array}
\right.
\label{eq:ParetoP}
\end{equation}
The discontinuity of $P_p(w)$ at $w=w_{\min}$ is worrisome, and most economists regard Pareto's observation as an approximation, at best.

\begin{figure}
\begin{center}
\includegraphics[bbllx=0,bblly=0,bburx=327,bbury=267,width=.40\textwidth]{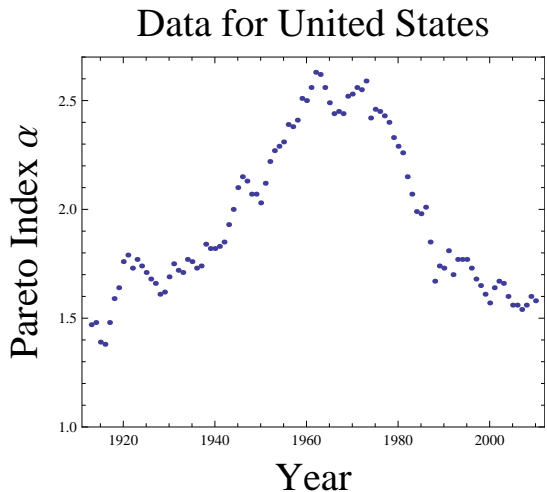}
\end{center}
\caption{{\bf The Pareto index of the US economy:}  Actual data for the last century, taken from~\cite{bib:TopIncomes}.}
\label{fig:ParetoIndex}
\end{figure}

Pareto's law is sometimes equated with the ``80-20 rule'' that asserts that 20\% of the population owns 80\% of the land.  In fact, this is implied by Pareto's law for $\alpha\approx 1.16$, but it does not, by itself, imply Pareto's law.  More generally, it is straightforward to show that Pareto's law can be made consistent with the observation that a fraction $f$ of the population has a fraction $1-f$ of the wealth if
\begin{equation}
\alpha = \frac{\log\left(\frac{1}{f}\right)}{\log\left(\frac{1-f}{f}\right)}.
\end{equation}
Note that the ``fair'' situation with $f=1/2$, in which half of the population owns half of the land, corresponds to $\alpha\rightarrow\infty$; the totally ``unfair'' situation, in which a vanishingly small fraction of the population owns all but a vanishingly small fraction of the land, corresponds to $\alpha\rightarrow 1$ from above.  Once again, this suggests that $\alpha> 1$.  The Pareto index for the economy of the United States over the last century~\cite{bib:TopIncomes} is shown in Fig.~\ref{fig:ParetoIndex}.

Although the details of the distribution of wealth in a society are controversial, the appearance of power laws in this context is widely accepted.  Power laws are often associated with self-similarity, which, in this context, is manifested by the following observation:  Denote the population with wealth between $w/2$ and $w$ by $N_-$, and that with wealth between $w$ and $2w$ by $N_+$.  Pareto's law holds~\footnote{Here we assume that $w/2>w_{\min}$ so that we are in the power-law regime.} if and only if $N_-/N_+=2^\alpha$, independent of $w$.  That is, the ratio of people within a factor of two poorer than $w$ to those within a factor of two wealthier than $w$ is independent of $w$.

Although Pareto's law has been known for more than a century, its microeconomic foundations are still a subject of active research.  In the mid-1990s, an innovative class of models, called {\it asset exchange models} (AEMs), were introduced for this purpose.  In this paper, we analyze a particularly interesting one of these, called the ``Yard-Sale Model'' (YSM), which can be used to predict the evolution of the PDF of wealth as a function of time.

Models of this general sort were considered in the economics literature by Angle~\cite{bib:Angle:1986wt} in the 1980s.  They were first introduced into the physics literature and analyzed using the techniques of mathematical physics by Ispolatov, Krapivsky and Redner~\cite{bib:Ispolatov:1998wj,bib:Krapivsky} in the late 1990s.  The need to impose conservation of agents, and the proper boundary condition at $w=0$ was clarified by Yakovenko~\cite{bib:Yakovenko:2009jt}.  The name ``Yard-Sale Model'' seems to have been coined by Hayes~\cite{bib:Hayes:2002ts} in a popular article published in 2002.  Since then, these models have been further studied by Chakraborti~\cite{bib:Chakraborti1,bib:Chakraborti2} and his coworkers, among others.

The YSM consists of $N$ economic agents, each endowed with only one quality, namely wealth $w$.  In the simplest version of this model, $w$ is a positive real number; that is, we do not allow agents to have negative net wealth.  This feature is enforced in the initial conditions and, as will become clear, the dynamics are designed to preserve it.

The simplest version of the YSM is a closed economic system.  The number of agents $N$ remains constant.  No wealth is imported, exported, generated or consumed, so the total wealth of the population $W$ also remains constant.  Wealth can only change hands, from one agent to another.  Therefore, agents can become wealthier only at the expense of other agents becoming poorer.

Neoclassical economics assumes that all agents are ``optimizing individuals,'' who are fully informed about their options, and make decisions based on their own financial best interests.  If this were really the case~\footnote{and if there were an absolute notion of value}, no net wealth would ever change hands.  Two agents might agree to exchange some wealth, but one or the other would refuse to enter into the transaction unless the wealth exchanged was equal.  Economists refer to this state of affairs as {\it perfect pricing}.  Under the assumption of perfect pricing, the exchange of wealth would leave $P(w,t)$ unaltered.

As described by Hayes~\cite{bib:Hayes:2002ts} and by Beinhocker~\cite{bib:Beinhocker}, perfect pricing does not happen in the real world.  Real people make mistakes, and some people are more clever about this than others.  It is unrealistic to expect that a person wishing to purchase a commodity will conduct an exhaustive search for the lowest price.  More often, they will search only long enough to find an acceptable price.  For these reasons, the wealth exchanged in transactions between agents may differ, and net wealth will change hands.  The YSM describes the dynamics of this process.

How much net wealth might be transferred from one agent to another in a given transaction?  Let us suppose that the amount transferred must be strictly less than the smaller of the wealths of the two agents participating in the transaction.  This will ensure that all agents maintain positive wealth.  In practice, we shall say that the net change of wealth is a fraction $\beta\in(0,1)$ of the wealth of the poorer of the two agents.

Once the net change of wealth has been determined, it remains to decide which agent loses it, and which agent wins it.  Of course, if one agent is assumed to be more clever than all the others, he/she is more likely to be the winner.  Such an assumption will have the effect of quickly concentrating wealth in the hands of the most clever agents.  To give our model economy every benefit of the doubt, therefore, let us assume that the agents are equally clever, so that either is equally likely to be the winner.

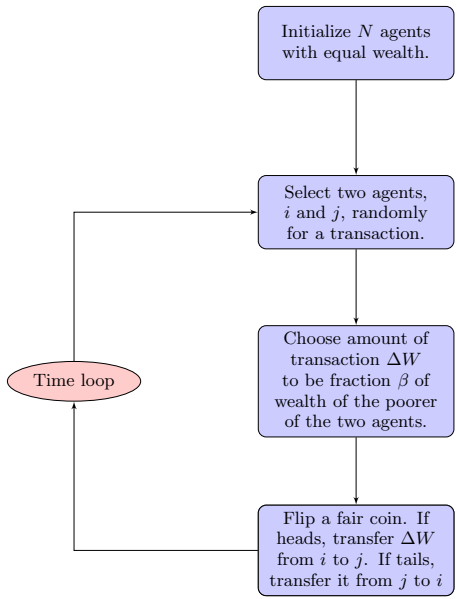
\begin{figure}
\begin{center}
\scalebox
{0.75} 
{
    \tikzstyle{block} = [rectangle, draw, fill=blue!20, 
    text width=10em, text centered, rounded corners, minimum height=4em]
    \tikzstyle{line} = [draw, -latex']
    \tikzstyle{cloud} = [draw, ellipse,fill=red!20, node distance=3cm,
    minimum height=2em]
\begin{tikzpicture}[node distance = 3.0cm, auto]
    \node [block] (init) {Initialize $N$ agents with equal wealth.};
    \node [block, below of=init] (select) {Select two agents, $i$ and $j$, randomly for a transaction.};
    \node [block, below of=select] (choose) {Choose amount of transaction $\Delta W$ to be fraction $\beta$ of wealth of the poorer of the two agents.};
    \node [cloud, left of=choose, node distance=5cm] (loop) {Time loop};
    \node [block, below of=choose] (decide) {Flip a fair coin.  If heads, transfer $\Delta W$ from $i$ to $j$.  If tails, transfer it from $j$ to $i$};
    \node [block, below of=choose] (flip) {Flip a fair coin.  If heads, transfer $\Delta W$ from $i$ to $j$.  If tails, transfer it from $j$ to $i$};
    \path [line] (init) -- (select);
    \path [line] (select) -- (choose);
    \path [line] (choose) -- (flip);
    \path [line] (flip) -| (loop);
    \path [line] (loop) |- (select);
\end{tikzpicture}
}
\end{center}
\caption{{\bf The time loop of the basic Yard Sale Model algorithm:}  As the algorithm proceeds, we keep track of the distribution of agent wealth versus time.}
\label{fig:ysm}
\end{figure}

These considerations lead to the simplest version of the YSM, which is described algorithmically in Fig.~\ref{fig:ysm}.  Note that the difference between the assumptions of neoclassical economics and those of this YSM could not be more stark.  In the former case, wealth transfer takes place between optimizing individuals, all of whom act in their best interests and try very hard not to make a mistake.  In the latter case, wealth is transferred only when somebody makes a mistake.

One of the principal results of this paper is that, under the assumption that $\beta$ is small, the equation governing the evolution of the PDF of the YSM is
\begin{equation}
\frac{\partial P}{\partial t} =
\frac{\partial^2}{\partial w^2}
\left[
\left(
\frac{w^2}{2}A + B
\right)
P
\right],
\end{equation}
where $A$ is given by Eq.~(\ref{eq:ParetoA0}), and $B$ is given by
\begin{equation}
B = \frac{1}{N}\int_0^w dw'\;P(w') \frac{{w'}^2}{2}.
\label{eq:PrincipalResult}
\end{equation}
Because AEMs are closed systems in which $N$ and $W$ are conserved, we expect that the dynamics described by this equation will drive the distribution $P(w,t)$ toward a steady state, dependent only on the values of $N$ and $W$, as $t\rightarrow\infty$.  Much of this paper is devoted to studying this limit.  We shall see that in the absence of mechanisms for wealth redistribution, this limit is a generalized function; with such mechanisms, it takes on a particular form that has some resemblance to the Pareto distribution, Eq.~(\ref{eq:ParetoP}).

\subsection{Outline of this paper}

In Sec.~\ref{sec:Boltzmann}, we consider the kinetics of the YSM, by relating the time rate of change of the one-agent distribution to an integral over the two-agent distribution for this model.  (A complete definition of multi-agent distributions is relegated to Appendix~\ref{sec:Klimontovich}.)  We derive this relation both by considering the outcome of a transaction between two agents, and from a master equation approach.  We then introduce the {\it random-agent approximation}, which is the analog of Boltzmann's famous {\it molecular chaos approximation}, to derive the analog of the Boltzmann equation for the YSM, Eq.~(\ref{eq:Boltzmann}).  We demonstrate that this equation conserves $N$ and $W$ for a closed economy; the details of this demonstration are relegated to Appendix~\ref{sec:conservationLawsBoltzmann}.  We believe this to be the first version of this transport equation which conserves both agents and wealth.

We give an exact solution to Eq.~(\ref{eq:Boltzmann}) that is non-normalizable, but we present numerical evidence that it is valid between lower and upper bounds of wealth.  We show that, in the long-time limit, these bounds tend to zero and infinity, respectively, as the result tends to a certain generalized function.  Appendix~\ref{sec:GenFun} contains a short detour through the theory of distributions in order to properly describe this generalized function.

In Sec.~\ref{sec:SmallTransaction}, we study a particularly interesting limit of the Boltzmann equation in which agents are allowed to stake only a small fraction of their wealth in any one transaction.  In this {\it small-transaction limit}, the Boltzmann equation reduces to the elegant partial integrodifferential equation, Eq.~(\ref{eq:PrincipalResult}), that admits to a simple analysis.  This equation is, we believe, entirely new in this context, and one of the principal new results of this paper.  We demonstrate that this equation admits the same conservation laws as the Boltzmann equation, and we present numerical simulations of its evolution.  We show that its time-asymptotic limit is the same generalized function described in Sec.~\ref{sec:Boltzmann}.  We conjecture that this evolution is approximately valid for many more complicated models of economies, such as the famous Sugarscape model of Epstein and Axtell~\cite{bib:Sugarscape}.

Finally, in Sec.~\ref{sec:Features} we show how the partial integrodifferential equation derived in Sec.~\ref{sec:SmallTransaction} can be extended to include effects such as inflation, production and taxation.  We present the dynamical equations with these features included, in the small-transaction limit.  We show that inflation and production simply result in a rescaling of the PDF of wealth.  By contrast, we show that taxation, by effecting a redistribution of wealth, leads to a steady state that has many features in common with that posited by Pareto~Eq.~(\ref{eq:ParetoP}).

\section{Boltzmann equation for density function}
\label{sec:Boltzmann}

\subsection{Agent density functions}
\label{ssec:adfs}

As described in Sec.~\ref{sec:intro}, the YSM supposes a population of $N$ agents, each with some wealth $w\in{\mathbb R}_+$.  The {\it one-agent density function} is the PDF of agents in wealth space at time $t$, and is denoted by $P(w,t)$.  It is defined so that the number of agents with wealth $w\in[a,b]$ at time $t$ is $\int_a^b dw\; P(w,t)$.  If the time variable is clear from the context, we usually omit it; for example, we might abbreviate $P(w,t)$ by $P(w)$.

In considering $P(w)$ to be a continuous function of $w$, we are already making an approximation.  Economic agents are inherently discrete.  If there are only $N$ of them, each with a particular value of wealth $w$, then the function $P(w)$ can have support at only $N$ discrete points.  Throughout this paper, however, we consider the domain of $P$ to be the continuum, and we regard $P$ as a smooth function of $w$.  The precise nature of this smoothing is discussed in detail in Appendix~\ref{sec:Klimontovich}.

The total number of agents is then given by the zeroth moment of $P$,
\begin{equation}
N = \int_0^\infty dw\; P(w,t),
\label{eq:N}
\end{equation}
and the total wealth of the agents is the first moment of $P$,
\begin{equation}
W = \int_0^\infty dw\; P(w,t) w.
\label{eq:W}
\end{equation}
The average wealth of an agent is then $W/N$.  In a closed economy, $N$ and $W$ are conserved quantities, independent of time.

For the mathematical description of the YSM, we shall also require the {\it two-agent density function}.  This is the PDF of pairs of agents in wealth space, and is denoted by $P(w,w',t)$.  It is defined so that the number of pairs of agents at time $t$, one having wealth $w\in[a,b]$ and the other having wealth  $w'\in[c,d]$, is $\int_a^b dw\int_c^d dw'\; P(w,w',t)$.  Again, if the time variable is clear from the context, we usually omit it; for example, we might abbreviate $P(w,w',t)$ by $P(w,w')$.

Once again, in regarding $P(w,w')$ as a smooth function, we are neglecting effects due to the discreteness of the agents, as explained in Appendix~\ref{sec:Klimontovich}.  For the two-agent distribution, the further approximation may be made that the agents are distributed independently, so that $P(w,w')$ is given by the product $P(w)P(w')$.  This is tantamount to neglecting inter-agent correlations, as is also explained in Appendix~\ref{sec:Klimontovich}.  This neglect is valid if (i) the initial conditions do not contain inter-agent correlations, and (ii) the dynamics do not generate inter-agent correlations.  The first of these conditions is something that we can simply demand; the validity of the second is less clear.  In Subsec.~\ref{ssec:raa} it will be argued that the second condition is valid for the YSM.  It should be emphasized that it may not be valid for more sophisticated models of wealth exchange, for which inter-agent correlations may play an important role.

\subsection{Pair interaction between agents}
\label{ssec:pair}

We now consider the problem of deriving a dynamical equation for the one-agent PDF, $P(w,t)$, of the YSM.  Because agents gain or lose wealth due only to transactions with other agents, we expect that the rate of change of the one-agent PDF depends on the two-agent PDF, and indeed this turns out to be the case.  We shall derive this result both by considering a transaction between a pair of agents, and then again by a master equation approach.

The scenario where one agent with wealth $\wl$ wins and one with wealth $\wl'$ loses is described by
\begin{eqnarray}
w &=& \wl + \alpha\min\left(\wl, \wl'\right)
\label{eq:microscopic1}\\
w' &=& \wl' - \alpha\min\left(\wl, \wl'\right),
\label{eq:microscopic2}
\end{eqnarray}
where $w>\wl$ is the new wealth of the winning agent, $w'<\wl'$ is the new wealth of the losing agent, and $\alpha\in[0,1)$ is the fraction of the smaller initial wealth that is exchanged in the transaction.  Equations~(\ref{eq:microscopic1}) and (\ref{eq:microscopic2}) describe a bijection on ${\mathbb R}_+^2$ with inverse
\begin{eqnarray}
\wl &=& w - \frac{\alpha}{1-\alpha}\min\left(\frac{1-\alpha}{1+\alpha} w, w'\right)
\label{eq:wl}\\
\wl' &=& w' + \frac{\alpha}{1-\alpha}\min\left(\frac{1-\alpha}{1+\alpha} w, w'\right).
\label{eq:wlp}
\end{eqnarray}
The Jacobian of this transformation is straightforwardly calculated to be
\begin{equation}
J(w,w')
=
\frac{\partial(\wl,\wl')}{\partial(w,w')}
=
\frac{1}{1+\alpha}\theta\left(w'-\frac{1-\alpha}{1+\alpha}w\right) +
\frac{1}{1-\alpha}\theta\left(\frac{1-\alpha}{1+\alpha}w-w'\right),
\label{eq:jac}
\end{equation}
where $\theta$ is the Heaviside function.

\subsection{Derivation of dynamic equation for density function}

If we suppose that a pair with wealth $(\wl,\wl')$ at time $t$ transforms into a pair with wealth $(w,w')$ at time $t+\dt$ with probability $\lambda \dt$, we must have
\begin{equation}
P(w,w',t+\dt) dw\;dw' = (\lambda \dt) P(\wl,\wl',t)d\wl\;d\wl' + (1-\lambda \dt)P(w,w',t)dw\;dw'
\end{equation}
or, employing the Jacobian, Eq.~(\ref{eq:jac}),
\begin{equation}
P(w,w',t+\dt) dw\;dw' = (\lambda \dt) P(\wl,\wl',t)J(w,w')dw\;dw' + (1-\lambda \dt)P(w,w',t)dw\;dw'.
\label{eq:diff1}
\end{equation}
If we cancel $dw$, integrate over $dw'$ and divide by $N$, we obtain
\begin{equation}
P(w,t+\dt) = \frac{\lambda \dt}{N} \int_0^\infty dw'\; P(\wl,\wl',t)J(w,w') + (1-\lambda \dt)P(w,t),
\label{eq:diff2}
\end{equation}
where it is understood that $\wl$ and $\wl'$ are functions of $w$ and $w'$ as given by Eqs.~(\ref{eq:wl}) and (\ref{eq:wlp}).  We subtract $P(w,t)$ from both sides, divide by $\dt$ and let $\dt\rightarrow 0$ to find
\begin{equation}
\frac{\partial P(w,t)}{\partial t} =
\frac{1}{N}\int_0^\infty dw'\; P(\wl,\wl',t)J(w,w') - P(w,t),
\end{equation}
where we have absorbed $\lambda$ into the time scale.  Finally, using Eqs.~(\ref{eq:wl}), (\ref{eq:wlp}), (\ref{eq:jac}) and (\ref{eq:projection}) and some straightforward calculation, we find the rate equation,
\begin{eqnarray}
\frac{\partial P(w,t)}{\partial t}
&=&
-\left[
P(w,t) -
\frac{1}{1+\alpha}
P\left(\frac{w}{1+\alpha},t\right)\right]
\nonumber\\
& &
+
\frac{1}{N}
\int_0^{\frac{w}{1+\alpha}} dw'\;
\left[
P\left(w - \alpha w',w',t\right)
-
\frac{1}{1+\alpha}
P\left(\frac{w}{1+\alpha},w',t\right)
\right].
\label{eq:halfBBGKY}
\end{eqnarray}

Equation~(\ref{eq:halfBBGKY}) is incomplete because we have not yet taken into account the equal possibility that the agent with wealth $\wl$ could lose, and that with wealth $\wl'$ could win.  The rate equation for that case can be derived exactly as above, but it is easy to see that the result differs from Eq.~(\ref{eq:halfBBGKY}) only by the substitution $\alpha\rightarrow-\alpha$.  Because agents win or lose with equal probability, the correct total rate is the average of the two, so the rate equation for the wealth distribution becomes
\begin{eqnarray}
\frac{\partial P(w,t)}{\partial t}
&=&
-\left[
P(w,t)
-
\frac{1}{2(1+\alpha)}
P\left(\frac{w}{1+\alpha},t\right)
-
\frac{1}{2(1-\alpha)}
P\left(\frac{w}{1-\alpha},t\right)
\right]
\nonumber\\
& &
+
\frac{1}{2N}
\int_0^{\frac{w}{1+\alpha}} dw'\;
\left[
P\left(w - \alpha w',w',t\right)
-
\frac{1}{1+\alpha}
P\left(\frac{w}{1+\alpha},w',t\right)
\right]
\nonumber\\
& &
+
\frac{1}{2N}
\int_0^{\frac{w}{1-\alpha}} dw'\;
\left[
P\left(w + \alpha w',w',t\right)
-
\frac{1}{1-\alpha}
P\left(\frac{w}{1-\alpha},w',t\right)
\right].
\label{eq:BBGKY}
\end{eqnarray}
Without this averaging of positive and negative rates, the resulting kinetic equation would not conserve the total wealth of the population, as we shall demonstrate in Appendix~\ref{sec:conservationLawsBoltzmann}.  In Subsec.~\ref{ssec:master}, we consider an alternative derivation of Eq.~(\ref{eq:BBGKY}).

We note that Eq.~(\ref{eq:BBGKY}) can be written in the form
\begin{eqnarray}
\frac{\partial P(w,t)}{\partial t}
&=&
\int_{-1}^{+1} d\beta\; \eta(\beta)
\left\{
-\left[
P(w,t)
-
\frac{1}{1+\beta}
P\left(\frac{w}{1+\beta},t\right)
\right]\right.
\nonumber\\
& &
+
\frac{1}{N}
\int_0^{\frac{w}{1+\beta}} dw'\;
\left.
\left[
P\left(w - \beta w',w',t\right)
-
\frac{1}{1+\beta}
P\left(\frac{w}{1+\beta},w',t\right)
\right]
\right\},
\label{eq:BBGKY2}
\end{eqnarray}
where $\eta$ is the PDF of the fraction $\alpha$ and is given by
\begin{equation}
\eta(\beta) := \frac{1}{2}\delta(\beta-\alpha) + \frac{1}{2}\delta(\beta+\alpha)
\label{eq:eta1}
\end{equation}
in the above example.  Note that we still regard $\alpha$ as confined to the interval $[0,1)$, but $\beta\in(-1,+1)$.  This form suggests that we could adopt a more general form for $\eta(\beta)$, as long as we retain the normalization $\int_{-1}^{+1}d\beta\;\eta(\beta) = 1$.  For example, by allowing the choice
\begin{equation}
\eta(\beta) =
\left\{
\begin{array}{ll}
\frac{1}{2\alpha} & \mbox{if $|\beta| < \alpha$}\\
0 & \mbox{otherwise,}
\end{array}
\right.
\label{eq:eta2}
\end{equation}
we model the situation in which the fraction of the poorer agent's wealth that is at stake is uniformly distributed in $[0,\alpha]$.  In any case, we demand that $\eta$ be an even function so that each agent has equal win and loss probabilities in each interaction.

\subsection{Master equation approach}
\label{ssec:master}

As has been pointed out by Ispolatov, Krapivsky and Redner~\cite{bib:Ispolatov:1998wj,bib:Krapivsky}, an excellent way to understand the origin of the terms in equations such as Eq.~(\ref{eq:BBGKY}) is to express them in the form of a master equation as follows
\begin{eqnarray}
\frac{\partial P(w,t)}{\partial t}
&=&
\frac{1}{N}
\int_0^{\infty} dw'
\int_0^{\infty} dw''\;
P(w'',w')
\left[
-\delta(w''-w)
\right.
\nonumber\\
& &
+\frac{1}{2}\theta\left(w-(1+\alpha)w'\right)
\delta\left(w''-w+\alpha w'\right)
\nonumber\\
& &
+\frac{1}{2}\theta\left((1+\alpha)w'-w\right)
\delta\left(w''(1+\alpha)-w\right)
\nonumber\\
& &
+\frac{1}{2}\theta\left(w-(1-\alpha)w'\right)
\delta\left(w''-w-\alpha w'\right)
\nonumber\\
& &
+\frac{1}{2}\left.\theta\left((1-\alpha)w'-w\right)
\delta\left(w''(1-\alpha)-w\right)
\right].
\label{eq:master}
\end{eqnarray}
We can think of the terms of Eq.~(\ref{eq:master}) as describing an agent with wealth $w''$ entering into a transaction with another agent with wealth $w'$.  The Dirac delta on the top line is a loss term; if $w''=w$, the transaction results in the loss of an agent with wealth $w$.  The four succeeding Dirac deltas are source terms, and may be justified as follows:
\begin{enumerate}
\item[(i)] In the first source term, the agent with wealth $w'' > w'$ wins wealth $\alpha w'$ from the agent with wealth $w'$, and becomes an agent with wealth $w = w'' + \alpha w' > (1+\alpha)w'$.
\item[(ii)] In the second source term, the agent with wealth $w''< w'$ wins wealth $\alpha w''$ from the agent with wealth $w'$, and becomes an agent with wealth $w = (1 + \alpha) w'' < (1+\alpha)w'$.
\item[(iii)] In the third source term, the agent with wealth $w'' > w'$ loses wealth $\alpha w'$ from the agent with wealth $w'$, and becomes an agent with wealth $w = w'' - \alpha w' > (1-\alpha)w'$.
\item[(iv)] In the fourth source term, the agent with wealth $w''< w'$ loses wealth $\alpha w''$ from the agent with wealth $w'$, and becomes an agent with wealth $w = (1 - \alpha) w'' < (1-\alpha)w'$.
\end{enumerate}
Note that each possibility (i) through (iv) supposes a win or a loss, and so each has a probability of one half.  Performing one or both integrals in each term of Eq.~(\ref{eq:master}) quickly yields Eq.~(\ref{eq:BBGKY}).

\subsection{Random-agent approximation and Boltzmann equation}
\label{ssec:raa}

Equation~(\ref{eq:BBGKY}) expresses the rate of change of the one-agent distribution in terms of the two-agent distribution.  We could proceed by writing an equation for the two-agent distribution, but it would involve the three-agent distribution.  This approach leads to an infinite hierarchy of equations, similar to the BBGKY hierarchy of statistical physics.

To truncate the hierarchy, we need to make an approximation.  Referring to Eq.~(\ref{eq:correlation}), we see that we can make the approximation of ignoring the correlation $C(w,w',t)$, so that the two-agent PDF is assumed to be a product of two one-agent PDFs.  In the context of kinetic theory, this is Boltzmann's famous {\it molecular chaos approximation}; in this context, we refer to it as the {\it random-agent approximation}.

The random-agent approximation assumes that two agents entering a transaction are uncorrelated.  It is of questionable validity.  We violate it every time we frequent the same grocery store, instead of choosing one randomly.  We will discuss the shortcomings of the random-agent approximation in Sec.~\ref{sec:conclude}.  For now we note that its application to Eq.~(\ref{eq:BBGKY2}) yields a self-contained dynamical equation for the one-agent PDF,
\begin{equation}
\boxed{
\mbox{
\addtolength{\linewidth}{-15\fboxsep}%
\addtolength{\linewidth}{-15\fboxrule}%
\begin{minipage}{\linewidth}
\begin{eqnarray}
\frac{\partial P(w,t)}{\partial t}
&=&
\int_{-1}^{+1} d\beta\; \eta(\beta)
\left\{
-\left[
P(w,t)
-
\frac{1}{1+\beta}
P\left(\frac{w}{1+\beta},t\right)
\right]\right.
\nonumber\\
& &
+
\frac{1}{N}
\int_0^{\frac{w}{1+\beta}} dw'\;
\left.
\left[
P\left(w - \beta w',t\right)
-
\frac{1}{1+\beta}
P\left(\frac{w}{1+\beta},t\right)
\right]P(w',t)
\right\}.
\nonumber\\
\nonumber
\end{eqnarray}
\end{minipage}
}}
\label{eq:Boltzmann}
\end{equation}
In Appendix~\ref{sec:conservationLawsBoltzmann}, we demonstrate that the quantities $N$ and $W$, defined in Eqs.~(\ref{eq:N}) and (\ref{eq:W}), are constants of the motion of Eq.~(\ref{eq:Boltzmann}).

Equation~(\ref{eq:Boltzmann}) is strongly reminiscent of Boltzmann's celebrated kinetic equation of statistical physics.  Certainly, the term with the integral over $w'$ on the right-hand side has the general appearance of an integral {\it collision operator} with quadratic nonlinearity.  We pursue this metaphor in Subsec.~\ref{ssec:comparison}.

\subsection{Comparison with statistical physics}
\label{ssec:comparison}

Boltzmann's kinetic equation of statistical physics is written for the one-particle PDF, $f(r,v,t)$, where $r$ denotes position and $v$ denotes velocity, and the evolution equation for this PDF has the form
\begin{equation}
\frac{\partial f(r,v,t)}{\partial t} =
-v\cdot\nabla f(r,v,t) + \Omega[f](r,v,t),
\label{eq:BoltzmannActual}
\end{equation}
where $\Omega[f](r,v,t)$ denotes a quadratically nonlinear integral collision operator whose detailed form is discussed at length in standard physics textbooks, and need not concern us here.

It is interesting to compare  the first term on the right of Eq.~(\ref{eq:BoltzmannActual}) to that of Eq.~(\ref{eq:BBGKY2}).  To address this, we rewrite this term in Eq.~(\ref{eq:BoltzmannActual}) as a finite difference
\begin{equation}
-v\cdot\nabla f(r,v,t) \approx -\frac{1}{\tau}\left[f(r,v,t) - f(r-v\tau,v,t)\right],
\end{equation}
where $\tau$ is small.  We note that both this term and the first term on the right-hand side of Eq.~(\ref{eq:BBGKY2}) involve the PDF minus a distortion of itself due to the action of a Lie group.  In Boltzmann's kinetic equation, the Lie group is that of galilean transformations, $r\rightarrow r-v\tau$.  In the Boltzmann equation that we have derived for the YSM economy, the Lie group is that of affine scalings $w\rightarrow w/(1+\beta)$.  Just as molecules move in physical space by addition of $-v\tau$, agents move in wealth space by multiplication by $1/(1+\beta)$.  Equation~(\ref{eq:BBGKY2}) may therefore be understood as a variety of Boltzmann equation that bears the same relation to the affine group as the physical Boltzmann equation bears to the Galilean group.

This observation strongly suggests that we should investigate the small-$\beta$ limit of Eq.~(\ref{eq:BBGKY2}) by considering PDFs $\eta(\beta)$ that have support only in the vicinity of the origin.  We shall examine this limit in Sec.~\ref{sec:SmallTransaction}.

\subsection{Solutions}
\label{ssec:solutions}
\subsubsection{Exact solutions}
\label{sssec:exact}

Ispolatov, Krapivsky and Redner~\cite{bib:Ispolatov:1998wj,bib:Krapivsky} investigated the Boltzmann equation obtained from applying the random agent approximation to Eq.~(\ref{eq:halfBBGKY}), and found that it admitted an exact solution proportional to $(wt)^{-1}$.  In fact, such solutions exist for the much more general Eq.~(\ref{eq:Boltzmann}), as can be verified by direct substitution.  Because Eq.~(\ref{eq:Boltzmann}) is manifestly invariant under time translation symmetry, these solutions can more generally be written as
\begin{equation}
P_{\mbox{\tiny exact}}(w,t) = \frac{C}{w(T+t)},
\label{eq:exact}
\end{equation}
where $T$ is an arbitrary constant, which should be positive to avoid a singularity at finite time, and where the constant $C$ is given by
\begin{equation}
C = 
\frac{N}{\int_{-1}^{+1} d\beta\;\eta(\beta)\ln\left(\frac{1}{1+\beta}\right)}.
\label{eq:exactC}
\end{equation}
The integral in the denominator in Eq.~(\ref{eq:exactC}) is a constant depending only on the choice of the symmetric function $\eta(\beta)$ used in the model.  For example, the choice of Eq.~(\ref{eq:eta1}) results in
\begin{equation}
C = 
\frac{N}{\ln\left(\frac{1}{\sqrt{1-\alpha^2}}\right)},
\end{equation}
and that of Eq.~(\ref{eq:eta2}) results in
\begin{equation}
C = 
\frac{N}{1 +
\frac{1}{2\alpha}
\ln\left(
\frac
{(1-\alpha)^{1-\alpha}}
{(1+\alpha)^{1+\alpha}}
\right)}.
\end{equation}

At first glance, the existence of such exact solutions might seem very useful.  Unfortunately, a solution proportional to $w^{-1}$ for all $w$ is not normalizable.  It has an infinite number of agents and an infinite total wealth.  That is, neither of the integrals in Eqs.~(\ref{eq:N}) and (\ref{eq:W}) are finite for these solutions.  The constant parameter $N$ in Eq.~(\ref{eq:exact}) is the same one that appears in Eq.~(\ref{eq:Boltzmann}), but it no longer has any connection with the  number of agents.

In spite of the fact that this solution is non-normalizable, we shall see that it is very useful in understanding the long-time behavior of solutions for $P(w,t)$.

\subsubsection{Simulations}

\begin{figure}
\begin{center}
\mbox{
\includegraphics[bbllx=0,bblly=0,bburx=360,bbury=263,width=2.5in]{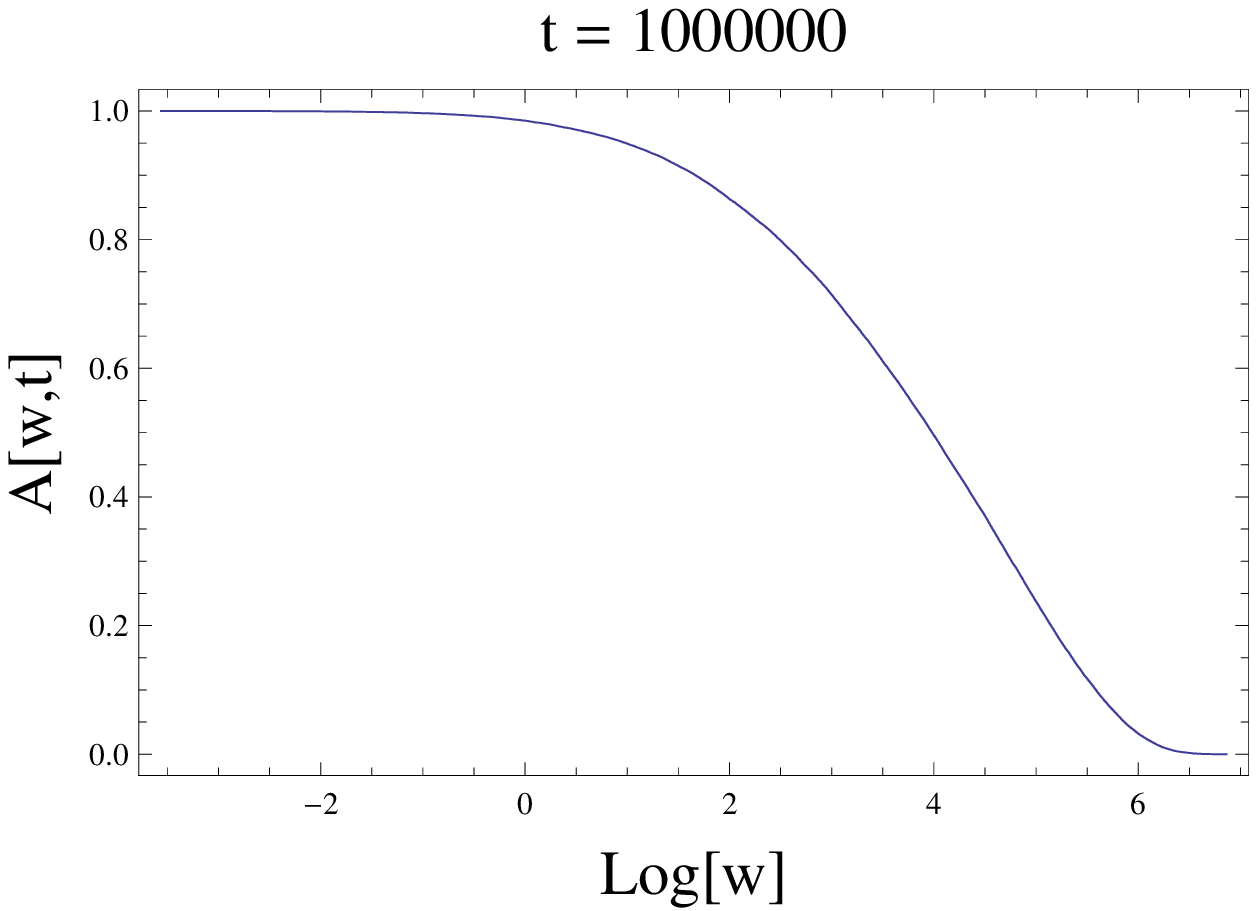}
\hspace{0.5in}
\includegraphics[bbllx=0,bblly=0,bburx=360,bbury=263,width=2.5in]{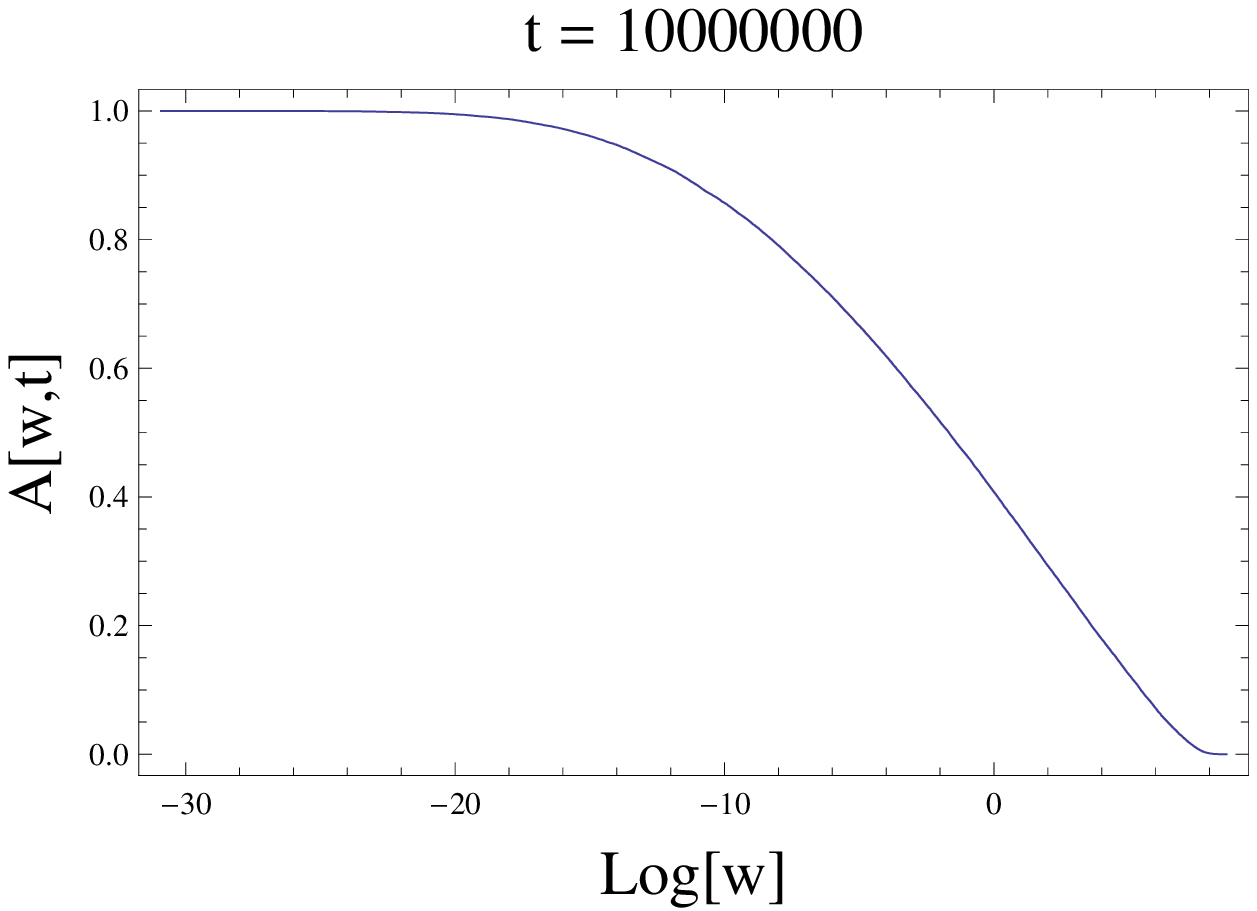}}\\
\mbox{
\includegraphics[bbllx=0,bblly=0,bburx=360,bbury=263,width=2.5in]{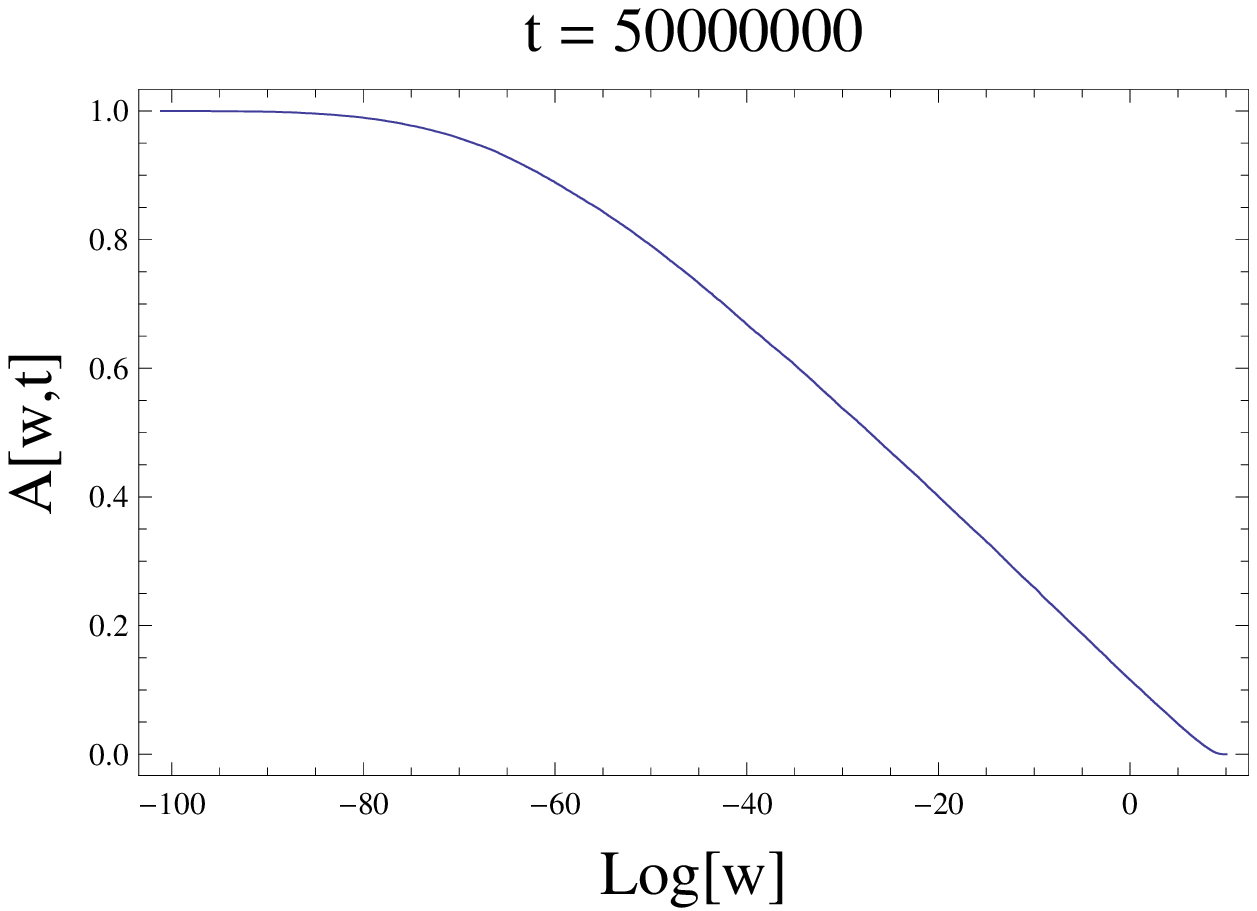}
\hspace{0.5in}
\includegraphics[bbllx=0,bblly=0,bburx=360,bbury=263,width=2.5in]{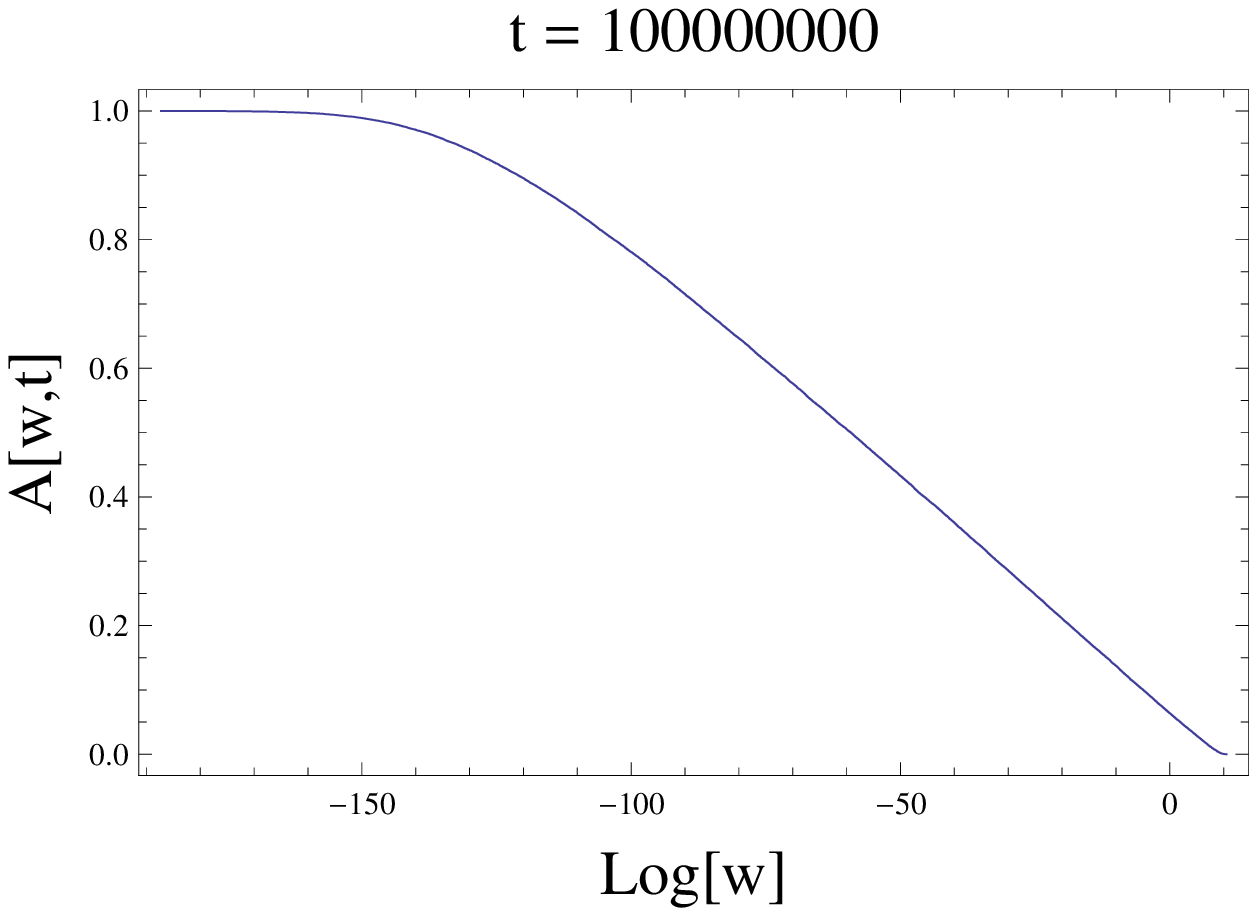}}\\
\mbox{
\includegraphics[bbllx=0,bblly=0,bburx=360,bbury=263,width=2.5in]{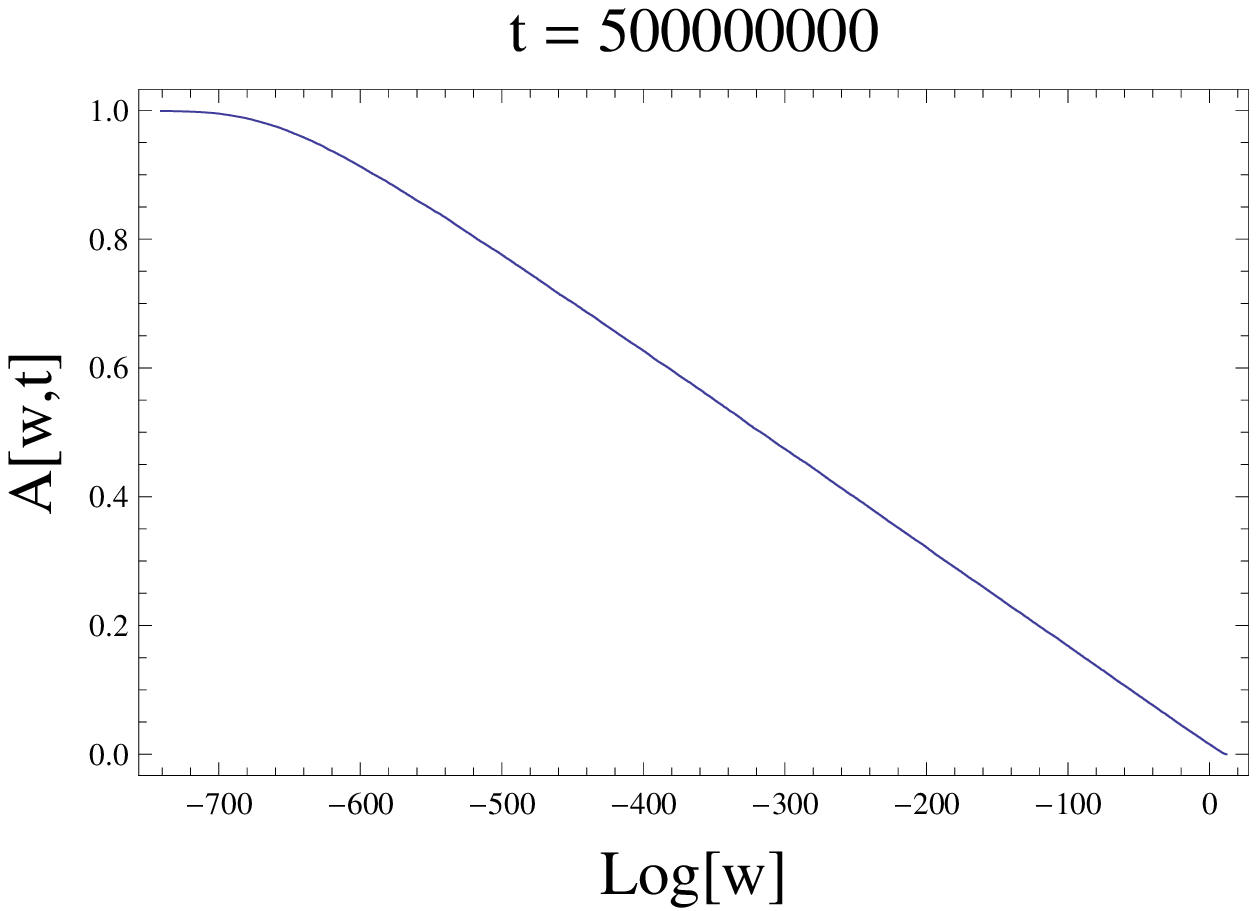}
\hspace{0.5in}
\includegraphics[bbllx=0,bblly=0,bburx=360,bbury=263,width=2.5in]{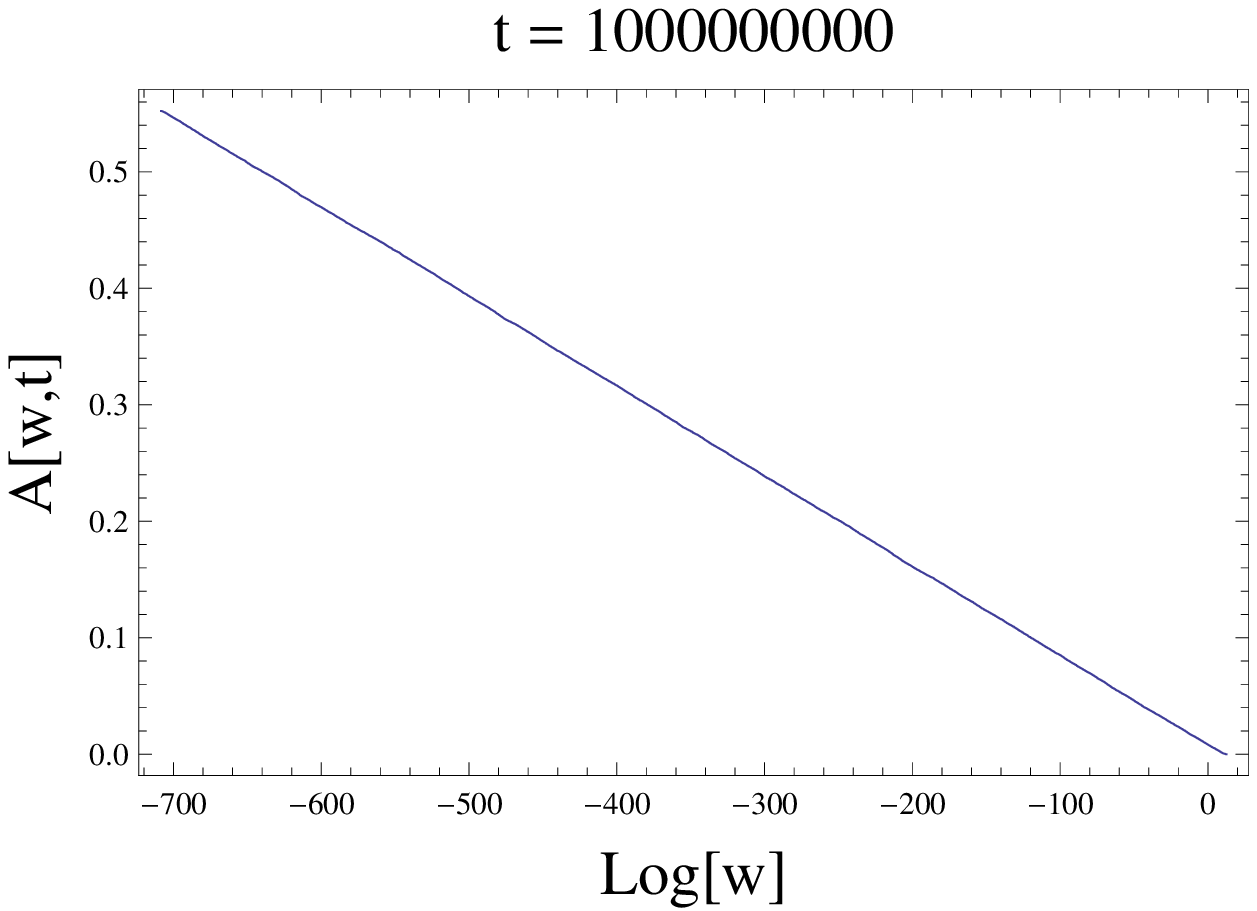}}
\end{center}
\caption{{\bf Log-linear Pareto plots of wealth distribution:} Taken from simulation for 50,000 agents, each with an initial allocation of 100 units of wealth and $\alpha = 0.25$.}
\label{fig:LogLinearPareto}
\end{figure}

We have performed simulations with populations of $N = 5\times 10^4$ agents, each given an initial allocation of 100 units of wealth, so that $W = 5\times 10^6$.  In these simulations, we took $\eta(\beta)$ to be of the form given in Eq.~(\ref{eq:eta1}), with $\alpha = 0.25$.  Using infinite-precision arithmetic, we ran the simulation for up to $10^9$ transactions and, following Pareto, we plotted the fraction of agents with wealth greater than $w$, namely
\begin{equation}
A(w,t):=\frac{1}{N}\int_w^\infty dw'\; P(w',t),
\label{eq:Adef}
\end{equation}
versus $w$.  These results are presented on log-linear plots for various times in Fig.~\ref{fig:LogLinearPareto}, in which three regimes are clearly visible.

\begin{itemize}
\item For sufficiently small values of $w$, we see $A(w,t)\approx 1$.  This indicates that $P(w,t)$ goes to zero for small enough $w$, so the lower limit of integration in Eq.~(\ref{eq:Adef}) may be replaced by zero.  It makes sense that $P(w,t)$ should vanish for sufficiently small $w$.  After all, at the beginning of the simulation, all the agents had 100 units of wealth.  Even an agent who lost in every one of his interactions would still have $100 (1-\alpha)^n > 0$ units of wealth remaining after $n$ transactions.  That said, it should be noted that the regime in which $A(w,t)\approx 1$ is restricted to extremely small values of $w$ indeed.  Remember that it is the logarithm of $w$ that is plotted on the abscissa in the graphs in Fig.~\ref{fig:LogLinearPareto}.  At time $t=10^8$, for example, note that the constant-$A$ regime is confined to $\ln w \lesssim -150$, or $w \lesssim e^{-150}$.  (This is why we used infinite-precision arithmetic in our simulations.)  We refer to this bound as $w_{\min}$, so this regime is defined by $w<w_{\min}$.
\item Figure~\ref{fig:LogLinearPareto} also suggests that $A(w,t)\approx 0$ for sufficiently large $w$.  This indicates that $P(w,t)$ also goes to zero for large enough $w$.  We refer to this bound as $w_{\max}$, so this regime is defined by $w>w_{\max}$.  Once again, this is reasonable, this time because there is a bound $W$ on the total wealth of the population.  Indeed, it may seem that it must be that $w_{\max}$ must be strictly less than $W$, but one must be careful about this.  It is true in our simulation because we have discrete agents; as a statement about Eq.~(\ref{eq:Boltzmann}), however, it is not true, because, as noted earlier, agent discreteness is lost in this representation, so we might well have a ``half an agent'' with wealth $2W$.  We will return to this point in more detail later.
\item For intermediate values of $w$, i.e., $w_{\min}<w<w_{\max}$, the curves in Fig.~(\ref{fig:LogLinearPareto}) fit well to straight lines with negative slope.  In this regime, we evidently have $A(w,t) \approx b(t) - a(t) \ln w$, and differentiating both sides with respect to $w$ yields $P(w,t) \approx \frac{a(t)}{w}$.  This looks remarkably like the exact solution presented earlier, but it is truncated for both low and high wealth.
\end{itemize}

\begin{figure}
\begin{center}
\includegraphics[bbllx=0,bblly=0,bburx=360,bbury=236,width=3.5in]{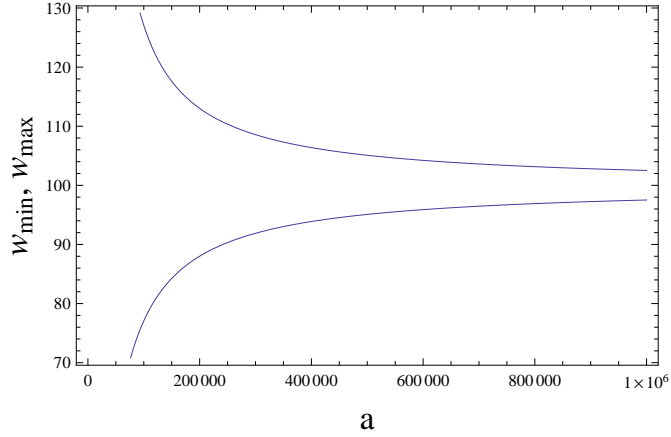}
\end{center}
\caption{{\bf Plot of $w_{\min}$ and $w_{\max}$ versus $a$:}  Computed from Eqs.~(\ref{eq:wMin}) and (\ref{eq:wMax}) for $N=50,000$ agents, and $W/N=100$ units of wealth.  The right-hand side of the plot corresponds to an egalitarian situation where most agents have wealth in the vicinity of $100$ units.  As time increases, $a$ decreases, leading to a wide range of wealth in the population, from the very poor to the very rich.}
\label{fig:wMinMax}
\end{figure}

The foregoing discussion suggests that, at any given time $t$, to a reasonable approximation, $P(w,t)$ has most of its support only on a finite interval, $[w_{\min}(t),w_{\max}(t)]$.  Thus our numerical results fit well to the approximate solution $P(w,t)\approx P_c(w,t)$, where
\begin{eqnarray}
P_c(w,t)
&:=&
\left\{
\begin{array}{ll}
\frac{a(t)}{w} & \mbox{for $w_{\min}(t) \leq w \leq w_{\max}(t)$}\\
0 & \mbox{otherwise,}
\end{array}
\right.
\label{eq:Pc}
\\
\noalign{\noindent{\mbox{from which it follows that}}}\nonumber\\
A_c(w,t)
&:=&
\left\{
\begin{array}{ll}
1 & \mbox{for $w \leq w_{\min}(t)$}\\
a(t)\log\left(\frac{w_{\max}(t)}{w}\right) & \mbox{for $w_{\min}(t)< w \leq w_{\max}(t)$}\\
0 & \mbox{for $w_{\max}(t) < w$},
\end{array}
\right.
\label{eq:Ac}
\end{eqnarray}
where the notation reflects the fact that $a$, $w_{\min}$ and $w_{\max}$ all depend on time $t$.  These quantities cannot all be independent, however, since they must satisfy
\begin{equation}
N = \int_0^\infty dw\; P_c(w,t) = a(t)\ln\left[\frac{w_{\max}(t)}{w_{\min}(t)}\right],
\end{equation}
and
\begin{equation}
W = \int_0^\infty dw\; P_c(w,t) w = a(t)\left[w_{\max}(t) - w_{\min}(t)\right].
\end{equation}
Solving these for $w_{\max}(t)$ and $w_{\min}(t)$, we find
\begin{eqnarray}
w_{\min} &=& \frac{W}{2a}\;\csch\left(\frac{N}{2a}\right)
\exp\left(-\frac{N}{2a}\right)
\label{eq:wMin}\\
\noalign{\noindent{\mbox{and}}}\nonumber\\
w_{\max} &=& \frac{W}{2a}\;\csch\left(\frac{N}{2a}\right)
\exp\left(+\frac{N}{2a}\right).
\label{eq:wMax}
\end{eqnarray}
Here we have suppressed the explicit dependences on time $t$, but the point is that the time dependence of $a$ determines those of $w_{\min}$ and of $w_{\max}$.  This dependence is plotted in Fig.~\ref{fig:wMinMax}, from which it is evident that large values of $a$ correspond to the egalitarian situation at early times, when everybody has approximately 100 units of wealth.  Small values of $a$ correspond to the situation at later times when there is a broad spectrum of wealth amongst the agents.  One might surmise, therefore, that $a(t)$ decreases in time, and we now turn our attention to measuring the rate at which it does so.

\begin{figure}
\begin{center}
\includegraphics[bbllx=0,bblly=0,bburx=360,bbury=230,width=3.5in]{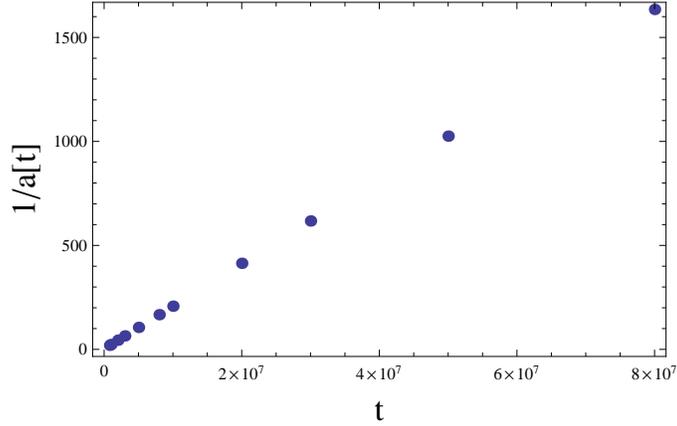}
\end{center}
\caption{{\bf Plot of $1/a(t)$ versus $t$:} Taken from numerical simulation by fitting to determine $w_{\min}(t)$ and solving Eq.~(\ref{eq:wMin}) numerically for $a(t)$, as described in text.}
\label{fig:aVersusT}
\end{figure}

Given the data in Fig.~\ref{fig:LogLinearPareto}, the easiest quantity for us to measure is $w_{\min}(t)$.  We fit the intermediate region of the curve -- the part with negative slope -- to a straight line, and determine where it intersects the horizontal line $A=1$.  Given $w_{\min}(t)$ calculated in this fashion, we solve Eq.~(\ref{eq:wMin}) numerically for $a(t)$, and plot $1/a(t)$ versus $t$.  The result, shown in Fig.~\ref{fig:aVersusT}, fits remarkably well to the straight line $1/a(t)\approx 3.93264 + 0.0000204046 t$ using a least-squares fit.  The slope is close to the value of $1/N = 0.00002$.  To within a multiplicative constant of order unity, we therefore conjecture the following approximate form for $a(t)$,
\begin{equation}
a(t) \approx \frac{N}{T+t},
\label{eq:at}
\end{equation}
where $T=N/a(0)$.

Combining Eqs.~(\ref{eq:Pc}) and (\ref{eq:at}), we see that, in the interval $[w_{\min}(t),w_{\max}(t)]$, our fit is very similar to the exact solution given in Eq.~(\ref{eq:exact}).  Outside this interval, however, $P_c(w)$ vanishes.  We emphasize that $P(w,t) = P_c(w,t)$ is merely a numerical fit, and it is not a (weak) solution of Eq.~(\ref{eq:Boltzmann}), as can be verified by direct substitution.  It can also be verified by noting that $A_c(w,t)$ has slope discontinuities at $w_{\min}(t)$ and $w_{\max}(t)$, whereas the numerically measured $A(w,t)$ in Fig.~\ref{fig:LogLinearPareto} seems smooth.  It is remarkable that this crude truncation of Eq.~(\ref{eq:exact}) does as well as it does in helping us understand the numerical results, but it does not explain them exactly.

Eqs.~(\ref{eq:Pc}) and (\ref{eq:Ac}) differ from the Pareto distribution of Eqs.~(\ref{eq:ParetoP}) and (\ref{eq:ParetoA}) in two significant ways.  First, there is an upper bound $w_{\max}$ as well as the lower bound $w_{\min}$.  Second, the effective Pareto index is $\alpha=0$ for this model.  The resulting distribution is normalizable only because of the imposition of the upper cutoff $w_{\max}$.

As mentioned earlier, measured values of the Pareto index seem to always be greater than one, as in Fig.~\ref{fig:ParetoIndex}, so it should be reemphasized that this is a very idealized model, and that we are not claiming that it models real economies.  More realistic models can be obtained by adding embellishments to this model, as will be described in Sec.~\ref{sec:Features}.  To pursue the metaphor with statistical thermodynamics, this model is the analog of the ideal gas law; no real economy obeys it, but it is such a useful idealization that it is worth careful study by anybody who endeavors to understand real economies.

\subsubsection{The long-time limit}

To what does the solution $P(w,t)$, or its approximation $P_c(w,t)$, converge in the limit of large $t$ or, equivalently, small $a$?  Because the process is a martingale, there cannot be a stationary solution that is a well behaved function, but we might expect that $P(w,t)$ and $P_c(w,t)$ converge to the same {\it generalized function} or {\it distribution}~\footnote{We shall use these two terms interchangeably.} as $t\rightarrow\infty$.  In Appendix~\ref{sec:GenFun}, we consider the nature of this generalized function, which we denote by $\zeta(w)$, and in what function space it exists.  The reader who is willing to accept at face value the statement, ``It converges to something that looks like a delta function at zero wealth, except that, somehow, it has a positive first moment and divergent higher moments,'' may skip the presentation in the Appendix without fear of losing the overall thread.

\section{A PDE for the Yard Sale Model density function}
\label{sec:SmallTransaction}
\subsection{The small-transaction limit}

In some circumstances, it is reasonable to assume that the largest fraction of an an agent's wealth that may be lost in one transaction is small.  Most sensible people, after all, do not stake large fractions of their wealth on a single transaction.  In that case, it is reasonable to expand the expression in curly brackets in Eq.~(\ref{eq:Boltzmann}) in a power series in $\beta$.  In doing so, we may note that this expression vanishes when $\beta=0$, so there is no constant term.  The next term of the power series, proportional to $\beta$, will contribute nothing when it is integrated alongside the even function $\eta(\beta)$.  Hence, the first term that contributes is that of order $\beta^2$.  The result, after some work, may be cast in the remarkably simple form
\begin{equation}
\boxed{
\frac{\partial P}{\partial t} =
\frac{\partial^2}{\partial w^2}
\left[
\left(
\frac{w^2}{2}A + B
\right)
P
\right],
}
\label{eq:smallBeta}
\end{equation}
where we have absorbed the constant factor $\int_{-1}^{+1} d\beta\; \eta(\beta) \beta^2$ into the unit of time $t$.  Here $A(w,t)$ is Pareto's function defined in Eq.~(\ref{eq:Adef}), and we have defined
\begin{equation}
B(w,t) := \frac{1}{N}\int_0^w dw'\; P(w',t) \frac{{w'}^2}{2}.
\label{eq:Bdef}
\end{equation}
Recall that $A(w,t)$ is non-increasing with $w$, with $A(0,t)=1$ and $\lim_{w\rightarrow\infty}A(w,t) = 0$.  By contrast, $B(w,t)$ is non-decreasing with $B(0,t)=0$, and $\lim_{w\rightarrow\infty}B(w,t)$ not necessarily finite.  Both $A(w,t)$ and $B(w,t)$ are functionals of $P$, so Eq.~(\ref{eq:smallBeta}) is nonlinear.  On the other hand, if $P$ is a solution, then so is $cP$ for any constant $c$, because $A$ and $B$ will be unchanged by this factor.

\subsection{Conservation laws in the small-transaction limit}

Before seeking solutions to Eq.~(\ref{eq:smallBeta}), we should check that we have retained the conservation laws in the limiting process.  Eq.~(\ref{eq:smallBeta}) is clearly in conservation form
\begin{equation}
\frac{\partial P}{\partial t} + \frac{\partial J_N}{\partial w} = 0,
\label{eq:consFormN}
\end{equation}
where we have defined the {\it flux of agents in wealth space},
\begin{equation}
J_N
=
-\frac{\partial}{\partial w}
\left[
\left(
\frac{w^2}{2}A + B
\right)
P
\right]
=
-\left(
\frac{w^2}{2}A + B
\right)
\frac{\partial P}{\partial w}
-
wAP.
\end{equation}
Because $J_N$ vanishes at the boundaries $w=0$ and $w\rightarrow\infty$, conservation of agents follows immediately by integration of Eq.~(\ref{eq:consFormN}) over all $w$.  Note that the quantity $\mu_N := \left(w^2 A/2 + B\right)P$ emerges as a kind of {\it chemical potential} for agents in wealth space, because its gradient drives the flux of agents, $J_N = -\partial\mu_ N/\partial w$.

Next note that we may write
\begin{equation}
0
=
w\frac{\partial P}{\partial t} + w\frac{\partial J_N}{\partial w}
=
\frac{\partial}{\partial t}\left(wP\right) + \frac{\partial}{\partial w}\left(wJ_N\right) - J_N
=
\frac{\partial}{\partial t}\left(wP\right) + \frac{\partial}{\partial w}\left(wJ_N + \mu_N\right),
\end{equation}
which is also in conservation form
\begin{equation}
\frac{\partial}{\partial t}\left(wP\right) + \frac{\partial J_W}{\partial w} = 0,
\label{eq:consFormW}
\end{equation}
where we have defined the {\it flux of wealth in wealth space},
\begin{equation}
J_W
=
wJ_N + \mu_N
=
-w\frac{\partial\mu_N}{\partial w} + \mu_N
=
-w\left(
\frac{w^2}{2}A + B
\right)
\frac{\partial P}{\partial w}
-
\left(
\frac{w^2}{2}A - B
\right) P.
\end{equation}
Because $J_W$ also vanishes at the boundaries $w=0$ and $w\rightarrow\infty$, conservation of wealth follows immediately by integration of Eq.~(\ref{eq:consFormW}) over all $w$.

\begin{figure}
\begin{center}
\includegraphics[bbllx=0,bblly=0,bburx=360,bbury=262,width=3.5in]{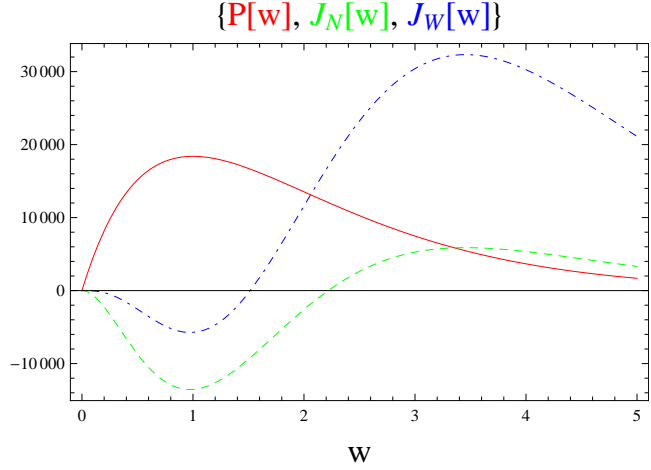}
\end{center}
\caption{{\bf Sample PDF and associated fluxes:}  A sample distribution $P(w)$ (in red, solid), the corresponding agent flux $J_N(w)$ (in green, dashed), and the corresponding wealth flux $J_W(w)$ (in blue, dot-dashed).}
\label{fig:fluxes}
\end{figure}

It is instructive to plot the agent flux and wealth flux as functions of $w$ for a sample distribution.  This plot is shown in Fig.~\ref{fig:fluxes} for the arbitrarily chosen distribution $P(w) = 50000 w e^{-w}$, which is normalized to 50,000 agents, and is plotted as a solid curve in red.  The corresponding $J_N(w)$ is plotted as a green dashed curve, and $J_W(w)$ as an blue dot-dashed curve.

Figure~\ref{fig:fluxes} makes evident that there is a threshold for agents in wealth space; the bulk of the agents below this threshold tend to move downward, while the elite above it tend to move upward.  Likewise, there is a different threshold for wealth; a minority of the wealth below this threshold tends to move downward, while the majority of wealth above it tends to move upward.  The agent threshold is on the tail of the distribution, significantly higher than the wealth threshold.  That is, a small fraction of the agents and a large fraction of the wealth move upward.  In this model, the rich become richer and the poor become poorer.

\subsection{Numerical simulations in the small-transaction limit}

\begin{figure}
\begin{center}
\mbox{
\includegraphics[bbllx=0,bblly=0,bburx=360,bbury=253,width=2.5in]{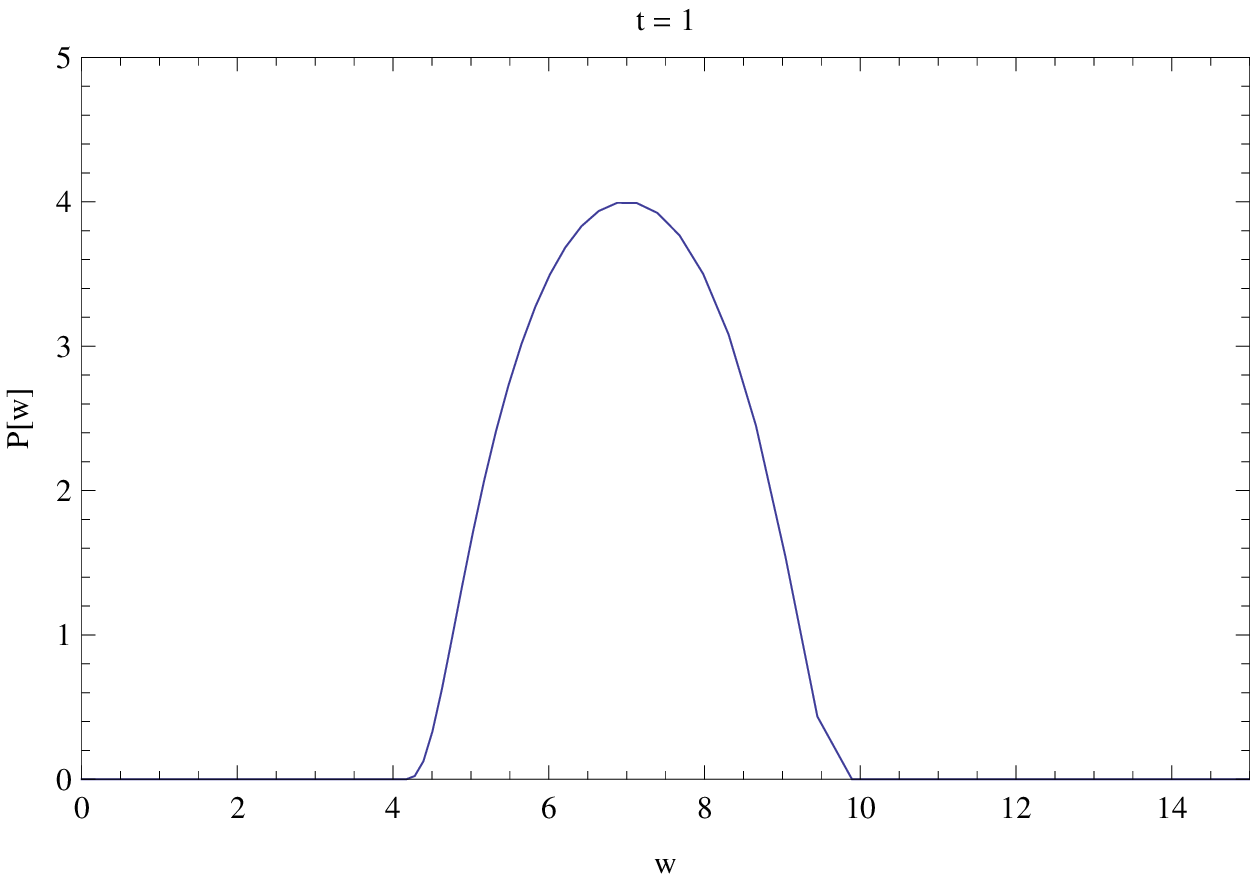}
\hspace{0.5in}
\includegraphics[bbllx=0,bblly=0,bburx=360,bbury=253,width=2.5in]{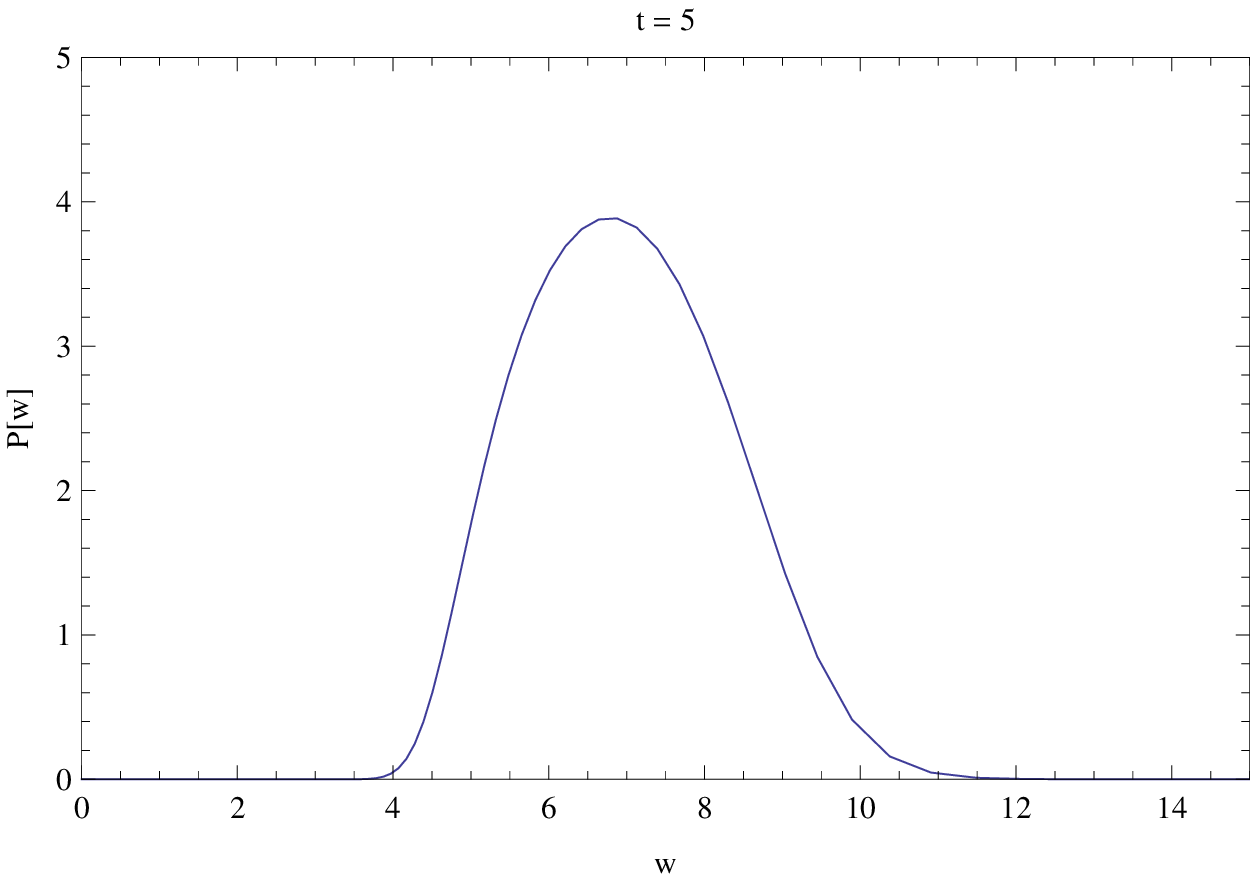}}\\
\mbox{
\includegraphics[bbllx=0,bblly=0,bburx=360,bbury=253,width=2.5in]{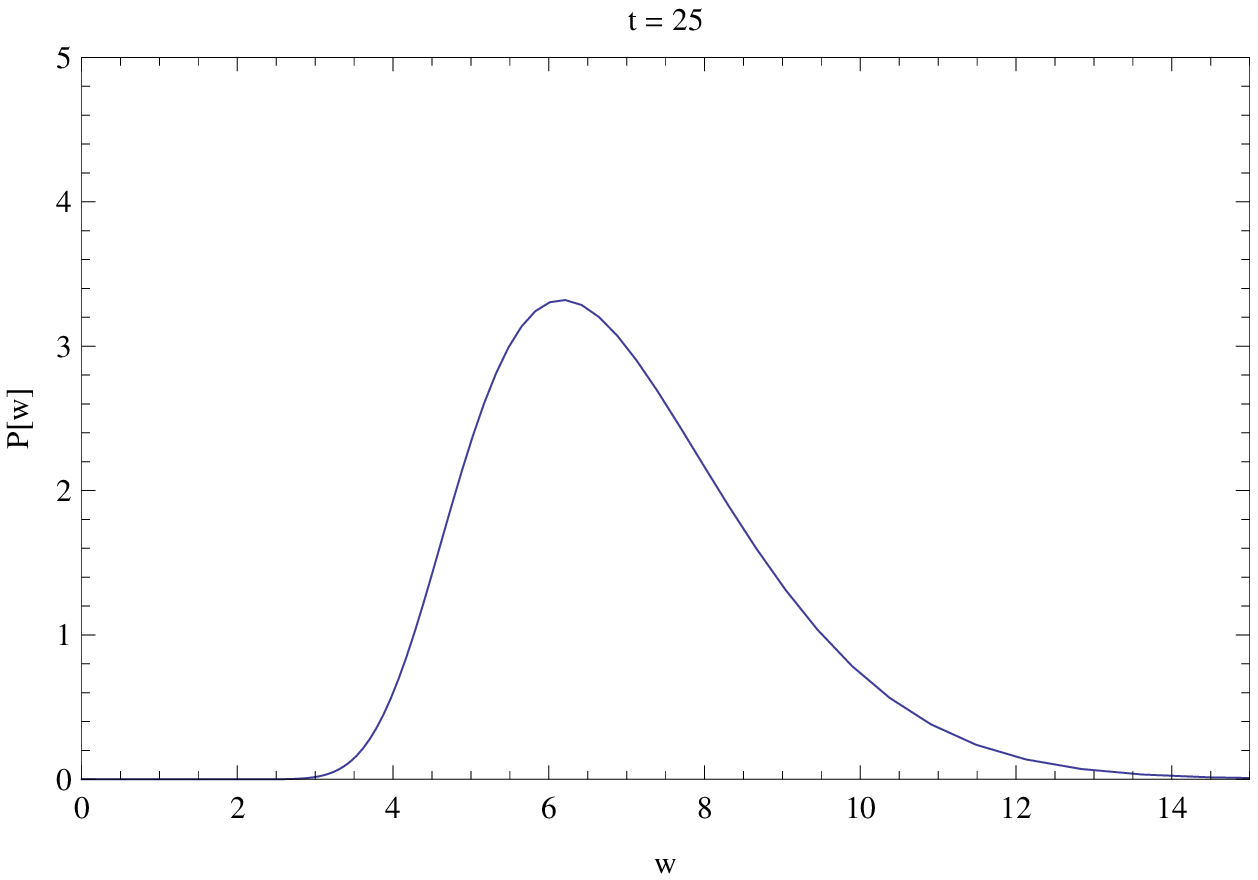}
\hspace{0.5in}
\includegraphics[bbllx=0,bblly=0,bburx=360,bbury=253,width=2.5in]{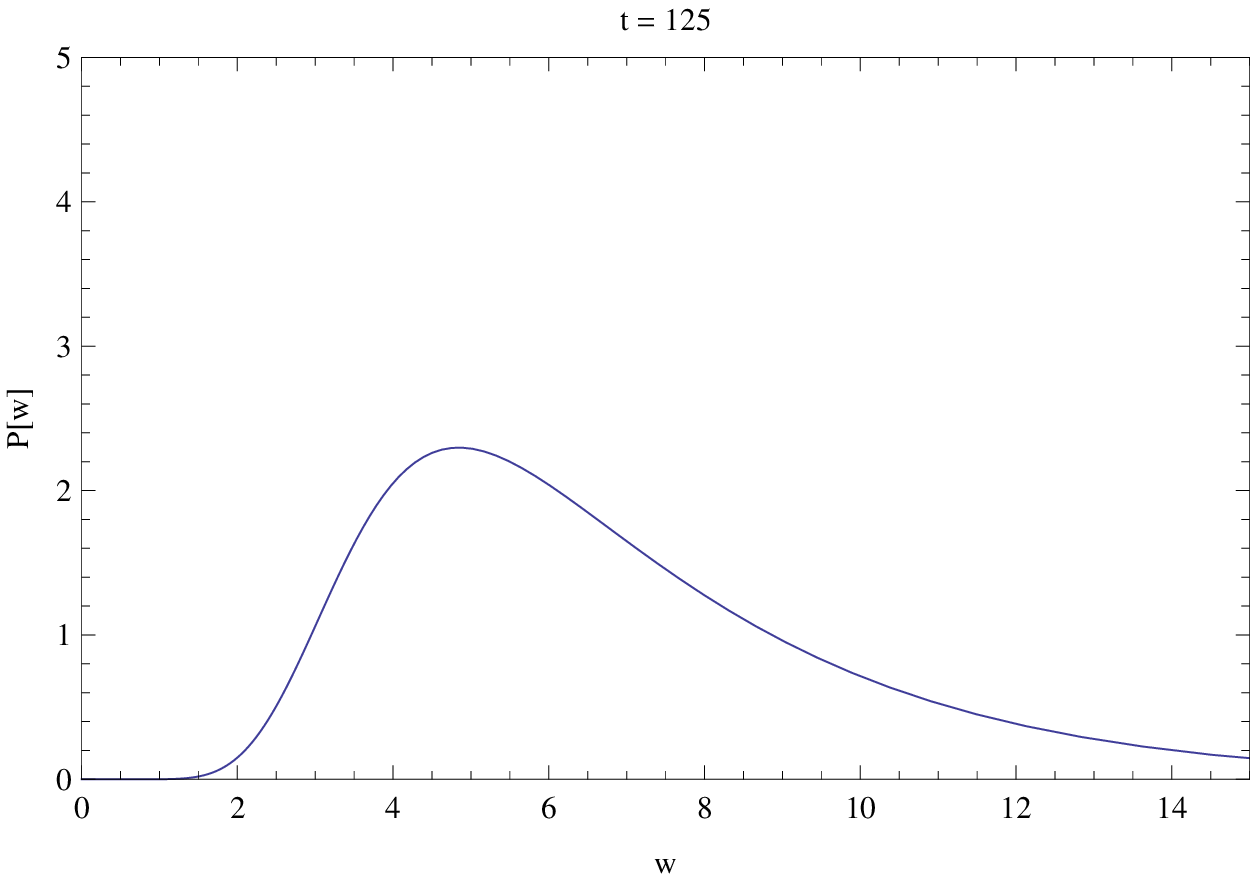}}\\
\mbox{
\includegraphics[bbllx=0,bblly=0,bburx=360,bbury=253,width=2.5in]{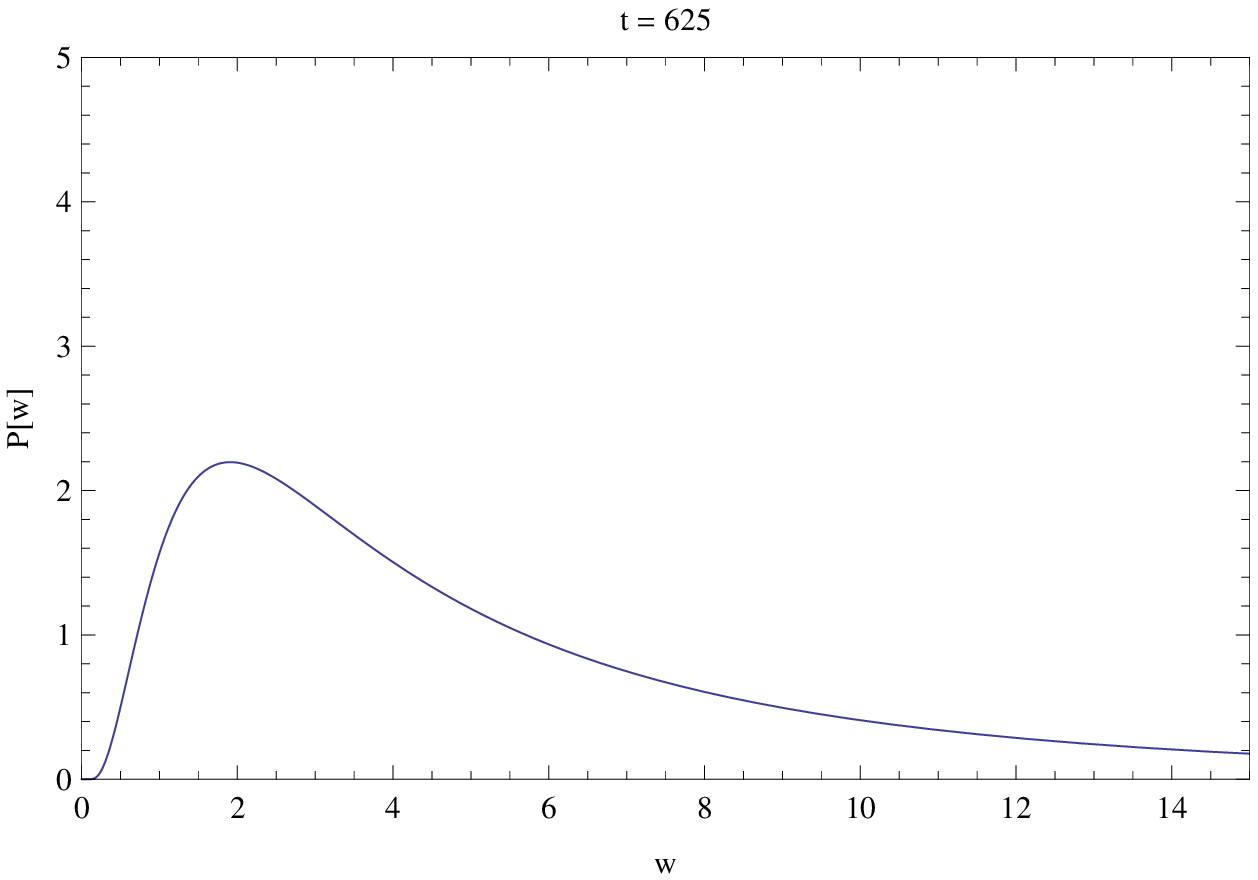}
\hspace{0.5in}
\includegraphics[bbllx=0,bblly=0,bburx=360,bbury=253,width=2.5in]{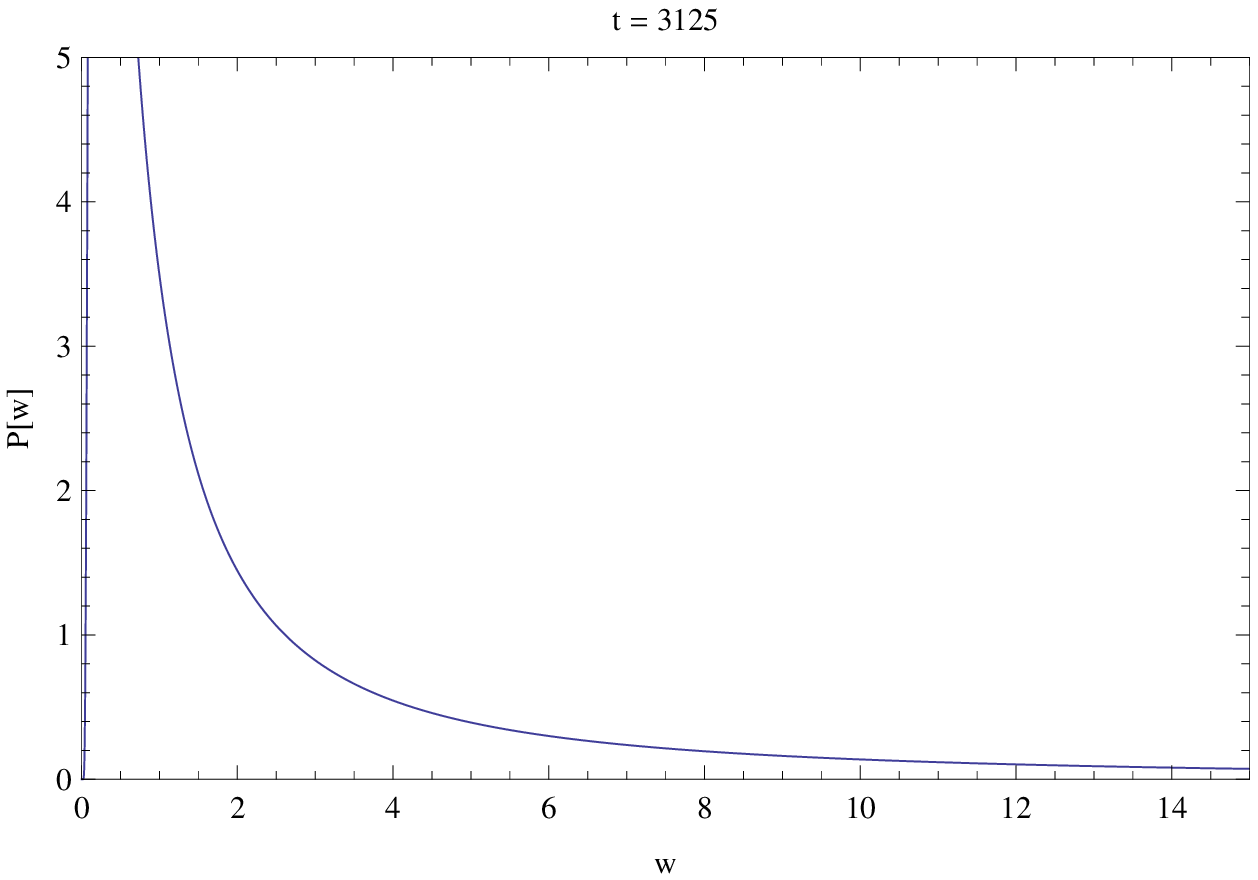}}
\end{center}
\caption{{\bf Numerical solution to Eq.~(\ref{eq:smallBeta}):} A finite-difference method was used to solve Eq.~(\ref{eq:smallBeta}) for $P(w,t)$, given the initial condition in Eq.~(\ref{eq:P0}).  The result clearly illustrates the approach to a curve proportional to $w^{-1}$, followed by the eventual approach to the singular distribution $\zeta(w)$.}
\label{fig:PDEsim}
\end{figure}

It is much more straightforward to simulate the PDE in Eq.~(\ref{eq:smallBeta}), with $A$ given by Eq.~(\ref{eq:Adef}) and $B$ given by Eq.~(\ref{eq:Bdef}), than it is to simulate Eq.~(\ref{eq:Boltzmann}).  We have done this using a finite-difference method for the arbitrarily chosen initial PDF,
\begin{equation}
P(w,0) \propto
\left\{
\begin{array}{ll}
\exp\left[-\frac{10}{(10-w)(w-4)}\right] & \mbox{for $4<w<10$}\\
0 & \mbox{otherwise,}
\end{array}
\right.
\label{eq:P0}
\end{equation}
which has support on $[4,10]$, and we plot the results in Fig.~\ref{fig:PDEsim}.  The results illustrate a fast evolution to a curve proportional to $w^{-1}$ in a bounded region, followed by the expansion of that region and concomitant reduction in magnitude of the curve, presumably approaching the singular function $\zeta(w)$ described in Appendix~\ref{sec:GenFun}.  At the end of the Appendix, we show that $\zeta(w)$ is a stationary state of Eq.~(\ref{eq:smallBeta}) in a weak sense.

\subsection{Discussion}

We have presented a Boltzmann equation for the YSM, and, in the small-transaction limit, we have shown that this reduces to a PDE.  Both are integrodifferential equations, though the second is easier to understand and simulate than the first.  Both agent-based numerical results from the Boltzmann equation, and a finite-difference simulation of the PDE reveal a strong tendency to drive increasing fractions of  wealth into the hands of a decreasing minority of agents.  In both cases, we conjecture that the time-asymptotic state of the system is a generalized function $\zeta(w)$ that has all of the $N$ agents condensed to zero wealth, while retaining a positive first moment $W$.

One might wonder if this approach to a singular state indicates that the model is lacking.  After all, even idealized agent-based models of microeconomics are much more complicated than the YSM.  As an example, consider the famous ``Sugarscape'' model of Epstein and Axtell~\cite{bib:Sugarscape}.  A condensed explanation of Sugarscape may be found in Beinhocker's book~\cite{bib:Beinhocker}, but even this explanation indicates that Sugarscape is vastly more complicated than our simple YSM.

In Sugarscape, agents have many features other than simply wealth.  For example, they have spatial location, and they can move about on a two-dimensional grid, searching for ``sugar'' and ``spice,'' and trading with other nearby agents.  They also have a built-in algorithm that controls their movements and actions based on their environment.  In the more sophisticated versions of the model, agents die for lack of sugar and breed when they have excess sugar.  There are also versions of the model in which the agents can sexually reproduce, with each parent passing along features of their algorithm to their offspring.

\begin{figure}
\begin{center}
\mbox{
\includegraphics[bbllx=22,bblly=304,bburx=583,bbury=483,width=3.0in]{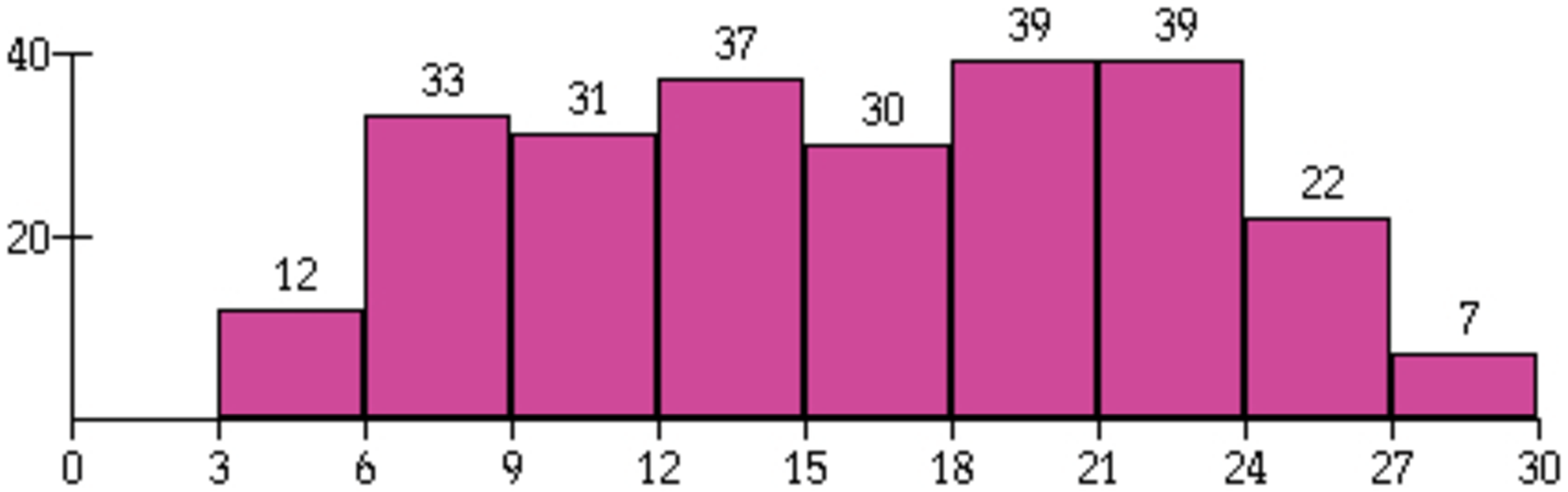}
}\\
\vspace{0.25in}
\mbox{
\includegraphics[bbllx=22,bblly=304,bburx=594,bbury=474,width=3.0in]{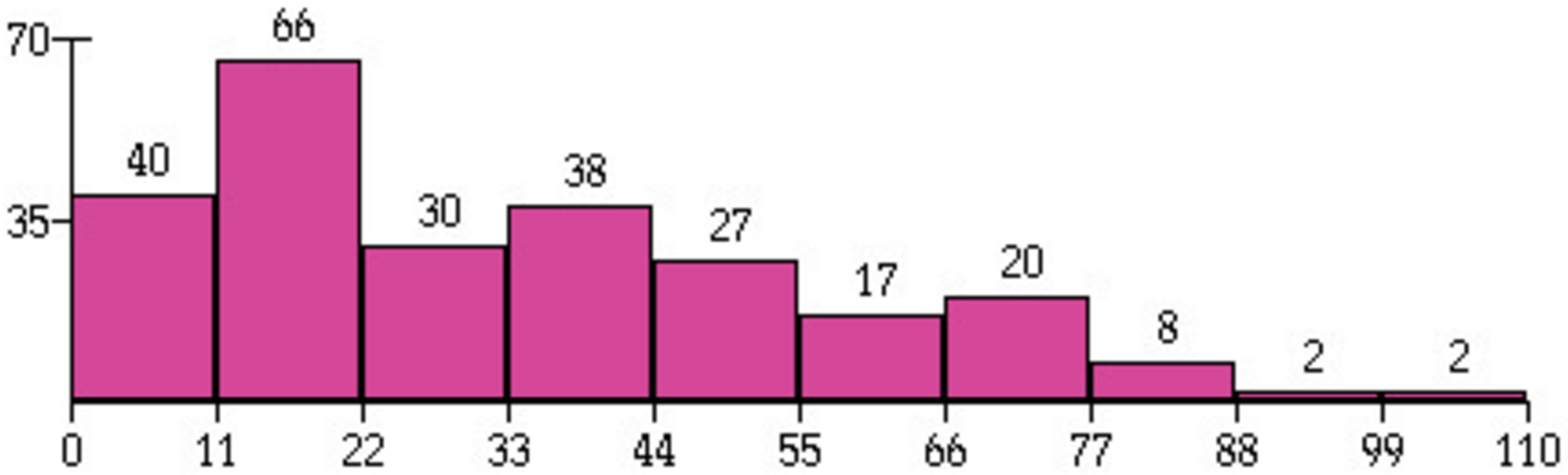}
}\\
\vspace{0.25in}
\mbox{
\includegraphics[bbllx=22,bblly=304,bburx=594,bbury=474,width=3.0in]{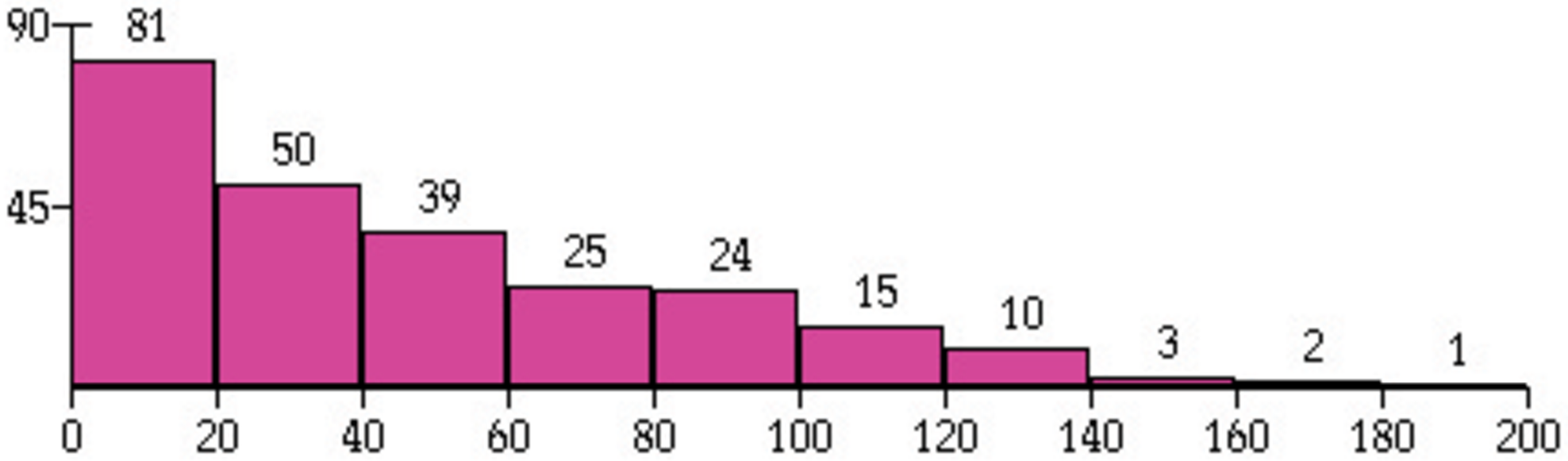}
}\\
\vspace{0.25in}
\mbox{
\hspace{-0.15in}
\includegraphics[bbllx=12,bblly=304,bburx=587,bbury=474,width=3.0in]{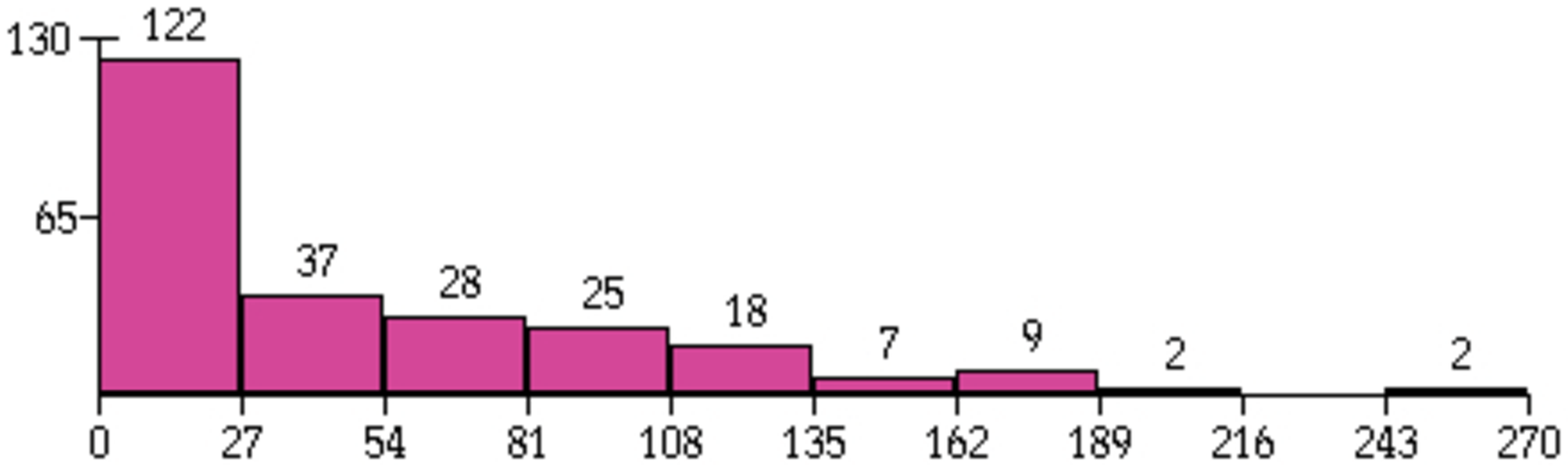}
}\\
\end{center}
\caption{{\bf Wealth distribution in ``Sugarscape'':} These plots are histograms of the number of agents versus wealth in Epstein and Axtell's  Sugarscape model.  Time runs downward from an arbitrary initial distribution in the top figure to something that looks remarkably like what is observed in the YSM.  (Figure taken from Epstein and Axtell~\cite{bib:Sugarscape} with permission.)}
\label{fig:SugarscapeWealth}
\end{figure}

Like us, Epstein and Axtell started the agents in Sugarscape with various initial distributions of wealth to see how these distributions would evolve, and they plotted their results versus time.  One of their time sequences is reproduced in Fig.~\ref{fig:SugarscapeWealth}.  Time runs downward in this figure.  In spite of all the complications present in Sugarscape, the evolution shown in Fig.~\ref{fig:SugarscapeWealth} is immediately familiar; indeed, the qualitative resemblance to Fig.~\ref{fig:PDEsim} is striking.  A least-squares fit on a log-log plot~\footnote{discarding histogram entries with zero agents} reveals that the penultimate plot in Fig.~\ref{fig:SugarscapeWealth} fits well to $w^{-1.36}$, and the last figure fits well to $w^{-1.24}$.  These correspond to Pareto $\alpha$ values of $0.36$ and $0.24$ -- not normalizable unless cutoffs are assumed.  These results are not so far removed from ours.

These observations suggest an Occam's Razor style argument that the YSM captures at least some of the essential features of Sugarscape, and there is no denying that the YSM is much simpler to understand and simulate.  Because I suspect that this paper will be read by economists as well as physicists, an additional transcultural cautionary word is warranted here.  Economists are naturally suspicious of the suggestion that correct macro predictions of a theory justify its microfoundations.  Nothing of the sort is being suggested here.  Sugarscape, while still very idealized, is far more realistic than the YSM.  In fact, it exhibits emergent phenomena, such as the growth of trade routes, that are not even defined in the YSM.

To a physicist, the fact that the YSM is able to explain some of the emergent phenomena of Sugarscape, such as power-law $P$ with $\alpha<1$, can only be regarded as a positive outcome.  Physicists have a long history of idealizations that have advanced human knowledge, from elliptic planetary orbits (Kepler), to arrows on a grid representing magnetic domains (Ising).  All of these idealizations are known to be unrealistic, and yet all of them have led to leaps in our understanding.  All we are suggesting here is that the YSM has a key place in the hierarchy of idealizations that constitute our understanding of real economic phenomena.

\section{Additional features}
\label{sec:Features}
\subsection{The importance of wealth redistribution}

Real economies seem to have Pareto exponents that are greater than one.  It is often claimed that $\alpha > 1$ is necessary in order for the Pareto PDF to be normalizable.  As we have seen, however, this argument is valid only if we assume no upper cutoff.  Real economies have discrete agents, so wealth can not concentrate beyond the extreme of one agent having all of it, and this in itself sets an upper cutoff.  As Pareto himself observed, there is also usually some social safety net for the poor, setting a lower cutoff.  With such cutoffs, there is nothing stopping the PDF between them from having a Pareto index less than unity, and this is precisely what we have found in both the YSM and Sugarscape models described above.

This naturally raises a question:  If normalizability is not the reason that $\alpha>1$ is observed in real economies, then what is the reason?  We suggest that real societies have wealth redistribution mechanisms that naturally increase $\alpha$.  It could be that real societies become politically unstable if $\alpha$ is too small.  Whatever the reason, most societies have taxation on wealth or income, and good governments use the revenues thereby generated to build infrastructure to improve the lives of all.

There are other mechanisms preventing the uncontrolled concentration of wealth.  Countries allow immigration to increase $N$, and they mine natural resources (among other things) to increase $W$.  Central banks can print currency.  Agents may make successful investments outside the country, thereby increasing their own wealth.  All of these features may impact the distribution of wealth.  We consider a few such features in the following subsections.

Recall that we have studied the YSM at two different levels of description, namely the Boltzmann equation in Sec.~\ref{sec:Boltzmann}, and the PDE to which it reduces in the small-transaction limit in Sec.~\ref{sec:SmallTransaction}.  We could introduce new features at either of these two levels of description.  In what follows, we continue to use the small-transaction limit because it is more elegant and tractable.  There is nothing preventing the use of a similar approach for the Boltzmann equation.

Suppose that a certain mechanism changes the wealth of an agent at a rate $f(w)$ that depends only on that agent's wealth $w$.  Then, to first order in $\Delta t$, we must have
\begin{equation}
P(w,t) dw = P(w+f(w)\Delta t,t+\Delta t) dw'.
\end{equation}
If we Taylor expand the right-hand side and retain terms only to first-order in $\Delta t$, we find
\begin{equation}
\frac{\partial P}{\partial t} +
\frac{\partial}{\partial w}\left(fP\right) = 0.
\label{eq:this2}
\end{equation}
Taking the zeroth moment of Eq.~(\ref{eq:this2}), we see that it conserves agents.  Taking the first moment, we see that Eq.~(\ref{eq:this2}) may not conserve wealth.  All of the examples that follow will conserve agents, so we shall use this general approach.

The observations in this section will be restricted to the derivation and exposition of appropriate dynamical equations.  Numerical modeling of economies with these extra features will be reported in a future paper~\cite{bib:BUGroup}.

\subsection{Inflation}
\label{ssec:inflation}

Suppose that all agents are able to loan their wealth to external borrowers who pay them an interest $\nu$ per unit time.  Then $f(w)=\nu w$, so if this mechanism were the only one present, the rate equation for the PDF would be
\begin{equation}
\frac{\partial P}{\partial t} +
\frac{\partial}{\partial w}\left(\nu w P\right) = 0.
\label{eq:this4}
\end{equation}
Once again, Eq.~(\ref{eq:this4}) conserves agents, but the total wealth of the society obeys
\begin{equation}
\frac{dW}{dt} = \nu W,
\end{equation}
demonstrating that $W$ grows exponentially in time, with time constant $\nu$, as expected.

If we suppose that this mechanism is present in addition to YSM wealth exchange, the full differential equation becomes
\begin{equation}
\frac{\partial P}{\partial t} +
\frac{\partial}{\partial w}\left(\nu w P\right) =
\frac{\partial^2}{\partial w^2}\left[
\left(
\frac{w^2}{2}A+B
\right)P
\right].
\end{equation}
Once again, because we have already demonstrated that the YSM terms on the right conserve both $N$ and $W$, this combined model has constant agent number $N$ and exponentially increasing wealth,
\begin{equation}
W = W_0 e^{\nu t}.
\end{equation}  

Because of the exponential increase of $W$, this model never reaches a stationary state, but we can rescale it by defining the new independent variables
\begin{eqnarray}
x &=& e^{-\nu t} w\\
\tau &=& t.
\end{eqnarray}
It follows that the derivatives with respect to the old variables are given by
\begin{eqnarray}
\frac{\partial}{\partial w} &=& e^{-\nu\tau}\frac{\partial}{\partial x}\\
\noalign{\noindent \mbox{and}}\nonumber\\
\frac{\partial}{\partial t} &=& -\nu x\frac{\partial}{\partial x} + \frac{\partial}{\partial\tau}.\\
\nonumber
\end{eqnarray}

The new dependent variable is then a new PDF, $Q$, such that $Q(x)\; dx = P(w)\; dw$.  From this, it follows that
\begin{equation}
Q = e^{\nu t} P,
\end{equation}
so that
\begin{eqnarray}
\int_0^\infty Q\; dx &=& \int_0^\infty P\; dw = N\\
\int_0^\infty Q x\; dx &=& e^{-\nu t} \int_0^\infty P w\; dw = e^{-\nu t} W = W_0.
\end{eqnarray}
Likewise, it follows that
\begin{eqnarray}
A &=& \frac{1}{N}\int_x^\infty Q\; dx\\
B &=& \frac{e^{2\nu t}}{N}\int_0^x Q\frac{x^2}{2}\; dx,
\end{eqnarray}
whence
\begin{equation}
\frac{w^2}{2}A+B = e^{2\nu\tau}\left(\frac{x^2}{2}\calA+\calB\right),
\end{equation}
where we have defined
\begin{eqnarray}
\calA &=& \frac{1}{N}\int_x^\infty Q\; dx\label{eq:calAdef}\\
\calB &=& \frac{1}{N}\int_0^x Q\frac{x^2}{2}\; dx.\label{eq:calBdef}
\end{eqnarray}

Assembling the above, we see that the differential equation for the new dependent variable $Q$ in terms of the new independent variables $x$ and $\tau$ is
\begin{equation}
\frac{\partial Q}{\partial \tau} =
\frac{\partial^2}{\partial x^2}\left[
\left(
\frac{x^2}{2}\calA+\calB
\right)Q
\right],
\label{eq:scaledInflation}
\end{equation}
where $\calA$ and $\calB$ are given in terms of $Q$ by Eqs.~(\ref{eq:calAdef}) and (\ref{eq:calBdef}).  Aside from renamed variables, however, Eqs.~(\ref{eq:scaledInflation}), (\ref{eq:calAdef}) and (\ref{eq:calBdef}) are absolutely identical in form to those for the YSM without inflation, Eqs.~(\ref{eq:smallBeta}), (\ref{eq:Adef}) and (\ref{eq:Bdef}).  Thus, the only effect of inflation in this closed economy is to change the yardstick by which wealth is measured, but the concentration of wealth predicted by the model persists.  In the long-time limit, the new dependent variable $Q$ approaches the generalized function $\zeta$ described earlier.

\subsection{Production}
\label{ssec:production}

Suppose that a society produces wealth $\xi$ per unit time, perhaps from an extraction industry of some sort, and that it divides the wealth thus produced evenly among its $N$ agents.  Then $f(w) = \xi/N$.  If this mechanism were the only one present, the rate equation for the PDF would be
\begin{equation}
\frac{\partial P}{\partial t} +
\frac{\partial}{\partial w}\left(\frac{\xi}{N}P\right) = 0.
\label{eq:this3}
\end{equation}
Eq.~(\ref{eq:this3}) is a one-sided wave equation with wave speed $\xi/N$.  As noted, it conserves the number of agents $N$.  Taking the first moment, however, we see that the total wealth of the society satisfies
\begin{equation}
\frac{dW}{dt} = \xi.
\end{equation}
In this model, therefore, $W$ grows linearly in time.

If we suppose that production occurs in addition to YSM wealth exchange, the full differential equation becomes
\begin{equation}
\frac{\partial P}{\partial t} +
\frac{\partial}{\partial w}\left(\frac{\xi}{N}P\right) =
\frac{\partial^2}{\partial w^2}\left[
\left(
\frac{w^2}{2}A+B
\right)P
\right].
\end{equation}
Because we have already demonstrated that the YSM terms on the right conserve both $N$ and $W$, this combined model will have constant agent number $N$ and linearly increasing wealth,
\begin{equation}
W=W_0+\xi t.
\end{equation}

Because of the linear increase of $W$, the model never reaches a stationary state, but we can  rescale it, as we did for the model with inflation.  There are a number of ways of going about this, but, for example, we could define the new independent variables~\footnote{We are going to use the same names for the new independent variables, $x$ and $\tau$, and for the new dependent variable, $Q$, that we used in the subsection on inflation, but obviously they will have different definitions in the context of production.},
\begin{eqnarray}
x &=& \frac{W_0}{W}w = \frac{W_0}{W_0+\xi t}w\\
\tau &=& t.
\end{eqnarray}
It follows that the derivatives with respect to the old variables are given by
\begin{eqnarray}
\frac{\partial}{\partial w} &=& \frac{W_0}{W_0+\xi\tau}\frac{\partial}{\partial x}\\
\noalign{\noindent \mbox{and}}\nonumber\\
\frac{\partial}{\partial t} &=& -\frac{\xi x}{W_0+\xi\tau}\frac{\partial}{\partial x} + \frac{\partial}{\partial\tau}.\\
\nonumber
\end{eqnarray}

The new dependent variable is then a new PDF, $Q$, such that $Q(x)\; dx = P(w)\; dw$.  From this, it follows that
\begin{equation}
P(w) = \frac{W_0}{W_0+\xi\tau}Q(x),
\end{equation}
so that
\begin{eqnarray}
\int_0^\infty Q\; dx &=& \int_0^\infty P\; dw = N\\
\int_0^\infty Q x\; dx &=& \frac{W_0}{W_0+\xi t}\int_0^\infty P w\; dw = \frac{W_0}{W}W = W_0.
\end{eqnarray}
Likewise, it follows that
\begin{eqnarray}
A &=& \frac{1}{N}\int_w^\infty P\; dw = \frac{1}{N}\int_x^\infty Q\; dx\\
B &=& \frac{1}{N}\int_0^w P\frac{w^2}{2}\; dw =
\left(\frac{W_0+\xi\tau}{W_0}\right)^2\frac{1}{N}\int_0^x Q\frac{x^2}{2}\; dx,
\end{eqnarray}
whence
\begin{equation}
\frac{w^2}{2}A+B = \left(\frac{W_0+\xi\tau}{W_0}\right)^2\left(\frac{x^2}{2}\calA+\calB\right),
\end{equation}
where we have defined
\begin{eqnarray}
\calA &=& \frac{1}{N}\int_x^\infty Q\; dx\label{eq:calAdef2}\\
\calB &=& \frac{1}{N}\int_0^x Q\frac{x^2}{2}\; dx.\label{eq:calBdef2}
\end{eqnarray}

Assembling the above, we see that the differential equation for the new dependent variable $Q$ in terms of the new independent variables $x$ and $\tau$ is
\begin{equation}
\frac{\partial Q}{\partial \tau} +
\frac{\xi\left[\left(\frac{W_0}{N} - x\right)\frac{\partial Q}{\partial x}-Q\right]}{W_0+\xi\tau} =
\frac{\partial^2}{\partial x^2}\left[
\left(
\frac{x^2}{2}\calA+\calB
\right)Q
\right].
\label{eq:scaledProduction0}
\end{equation}
Assuming that the various derivatives are well behaved, and taking the limit as $\tau\rightarrow\infty$, we find
\begin{equation}
\frac{\partial Q}{\partial \tau} =
\frac{\partial^2}{\partial x^2}\left[
\left(
\frac{x^2}{2}\calA+\calB
\right)Q
\right],
\label{eq:scaledProduction}
\end{equation}
where $\calA$ and $\calB$ are given in terms of $Q$ by Eqs.~(\ref{eq:calAdef2}) and (\ref{eq:calBdef2}).  Once again, aside from renamed variables, however, Eqs.~(\ref{eq:scaledProduction}), (\ref{eq:calAdef2}) and (\ref{eq:calBdef2}) are absolutely identical in form to those for the YSM without production, Eqs.~(\ref{eq:smallBeta}), (\ref{eq:Adef}) and (\ref{eq:Bdef}).  Thus, as with inflation, the only effect of production in this closed economy is to change the yardstick by which wealth is measured, but the concentration of wealth predicted by the model persists.  In the long-time limit, the new dependent variable $Q$ approaches the generalized function $\zeta$ described earlier.

\subsection{Taxation}
\label{ssec:taxation}

The importance of wealth redistribution in models of this sort has been emphasized by Toscani and his coworkers~\cite{bib:Toscani1,bib:Toscani2,bib:Toscani3,bib:Toscani4}.  To incorporate this effect in our model, let us suppose that all agents are assessed a wealth tax of $\chi$ percent per unit time.  The amount of tax paid by an agent with wealth $w$ is $\chi w$.  Integrating this over the distribution, we see that the total tax taken from the society is $\chi W$.  If we suppose that this total tax revenue is divided evenly and redistributed amongst the $N$ agents, we find that $f(w) = -\chi w + \chi W/N$.  If this mechanism were the only one present, the rate equation for the PDF becomes
\begin{equation}
\frac{\partial P}{\partial t} +
\frac{\partial}{\partial w}\left[\chi \left(\frac{W}{N}-w\right) P\right] = 0.
\label{eq:this5}
\end{equation}
Eq.~(\ref{eq:this5}) conserves both $N$ and $W$.  Because it continually redistributes wealth, it is not surprising that it admits the generalized stationary solution $P(w) = N\delta(w-W/N)$, in a weak sense, as is readily verified.

If we suppose that taxation is present in addition to YSM wealth exchange, the full differential equation is
\begin{equation}
\frac{\partial P}{\partial t} +
\frac{\partial}{\partial w}\left[\chi \left(\frac{W}{N}-w\right) P\right] =
\frac{\partial^2}{\partial w^2}\left[
\left(
\frac{w^2}{2}A+B
\right)P
\right].
\label{eq:taxation}
\end{equation}
This combined model will conserve both $N$ and $W$, and is interesting in that the terms on the left-hand side drive $P$ towards an equitable distribution of wealth, while those on the right-hand side drive $\alpha$ to zero.  We might hope that together they would lead to power-law solutions with intermediate values of the Pareto index, closer to those observed in real economies, but it is straightforward to verify that a simple power law will not work, even with upper and lower cutoffs.

\begin{figure}
\begin{center}
\includegraphics[bbllx=0,bblly=0,bburx=360,bbury=249,width=3.0in]{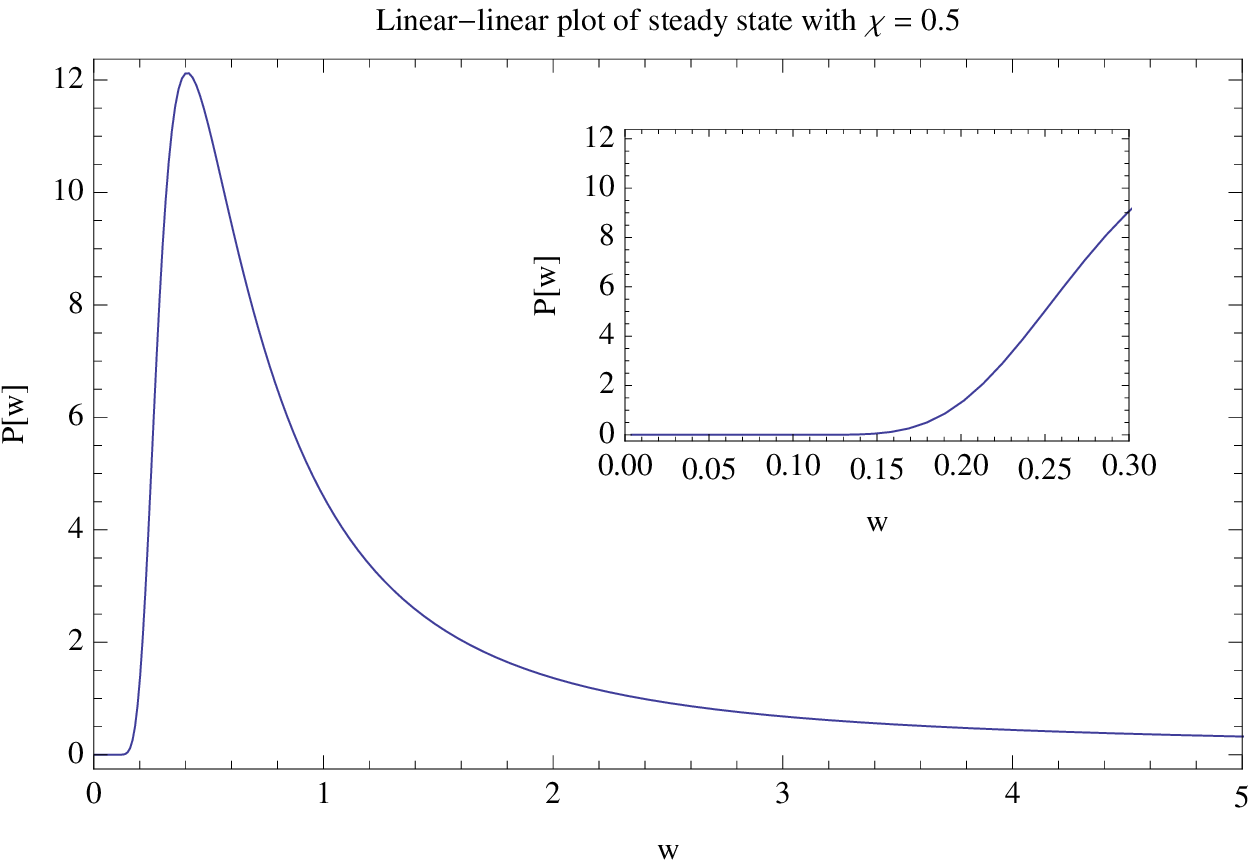}
\hspace{0.1in}
\includegraphics[bbllx=0,bblly=0,bburx=360,bbury=245,width=3.0in]{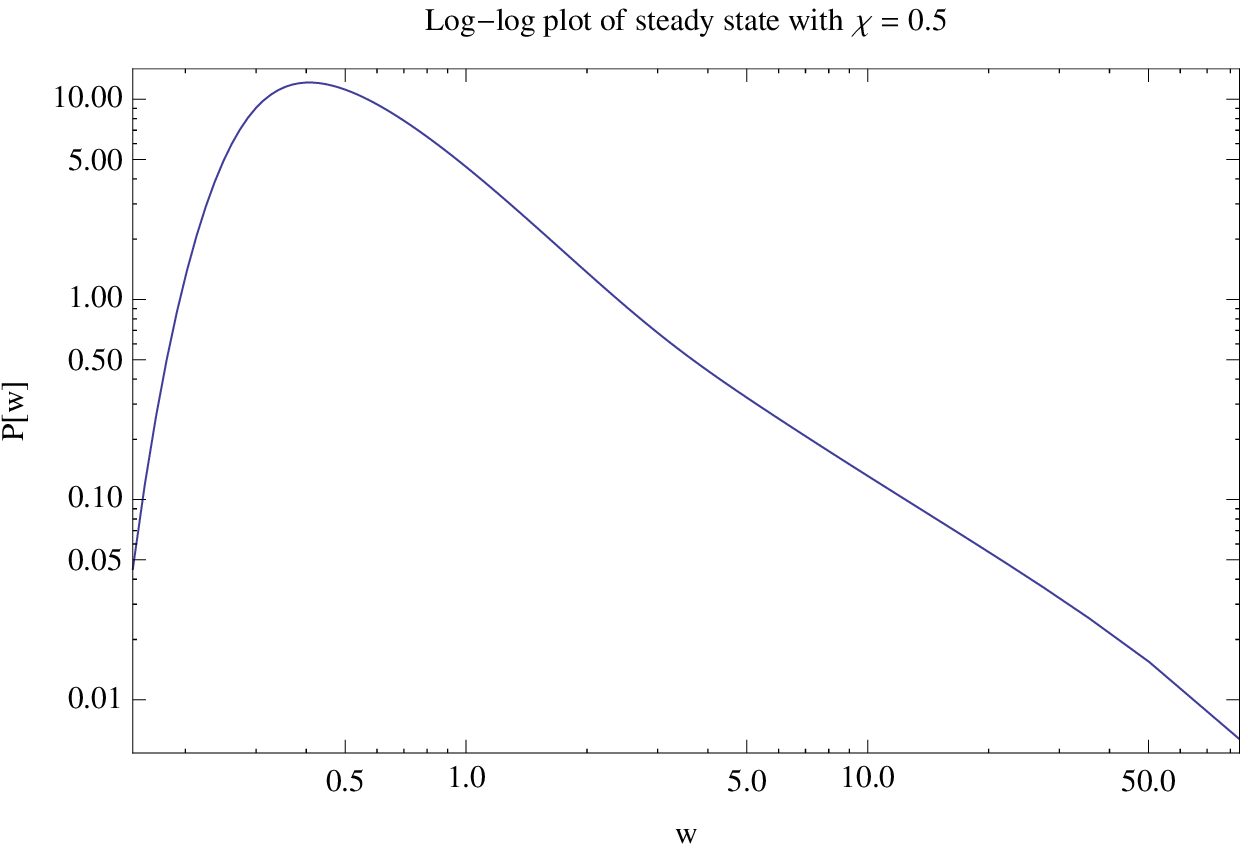}\\
\includegraphics[bbllx=0,bblly=0,bburx=360,bbury=249,width=3.0in]{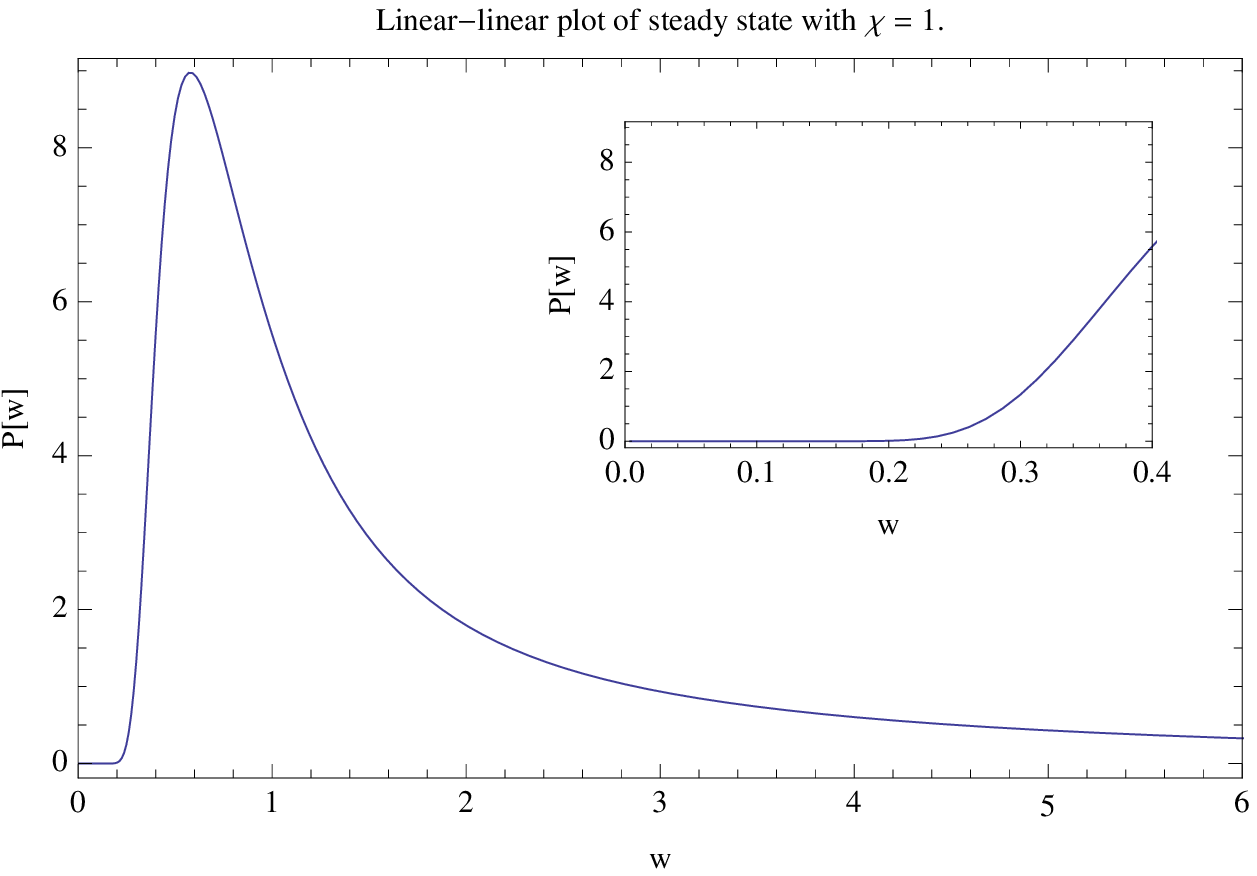}
\hspace{0.1in}
\includegraphics[bbllx=0,bblly=0,bburx=360,bbury=245,width=3.0in]{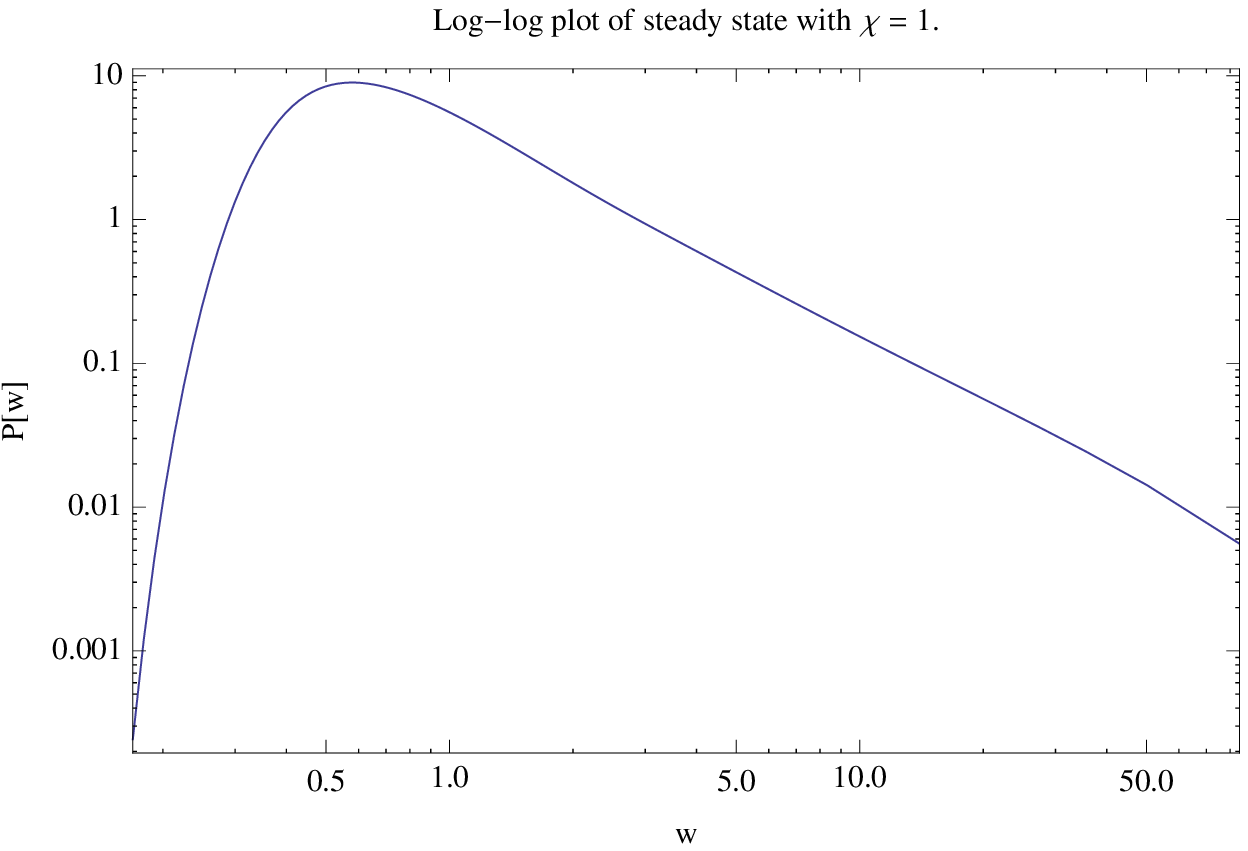}\\
\includegraphics[bbllx=0,bblly=0,bburx=360,bbury=249,width=3.0in]{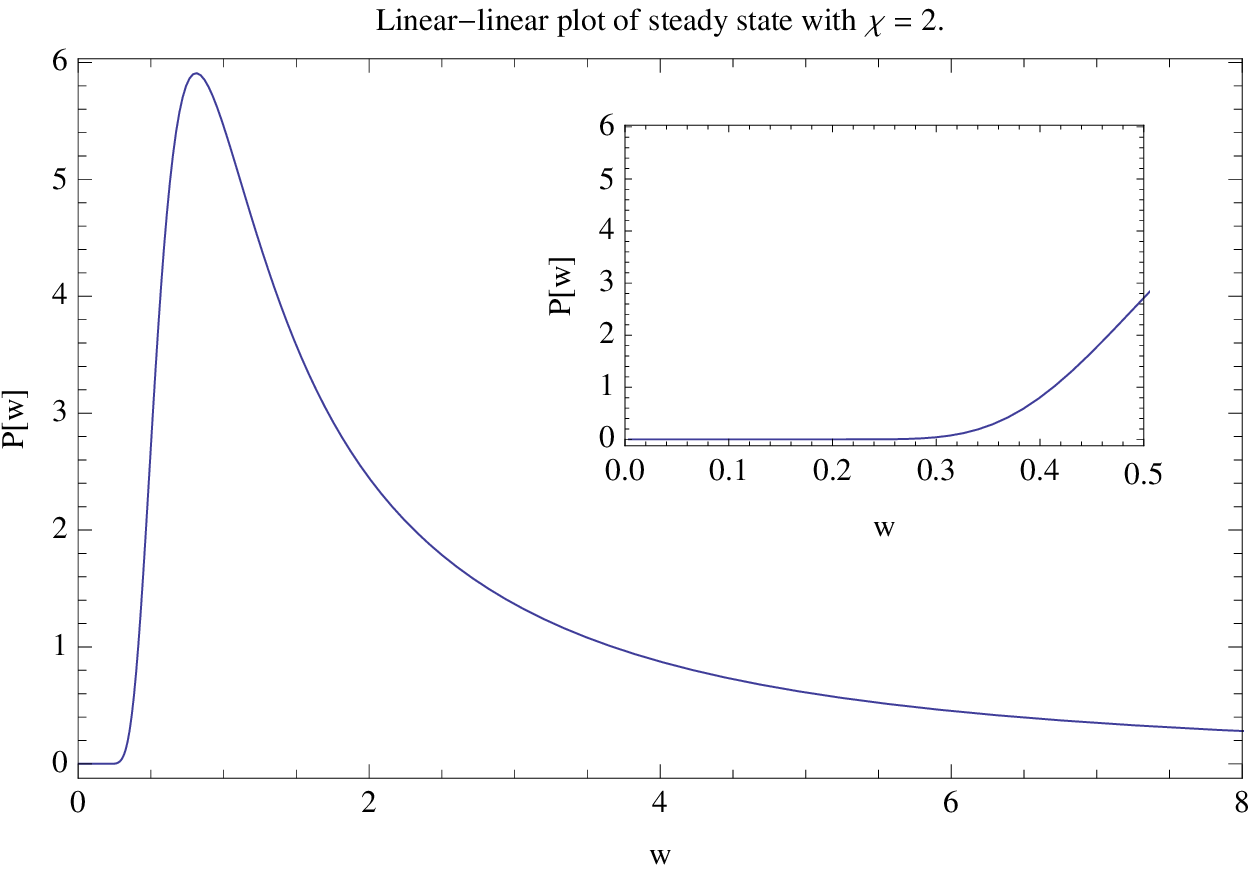}
\hspace{0.1in}
\includegraphics[bbllx=0,bblly=0,bburx=360,bbury=245,width=3.0in]{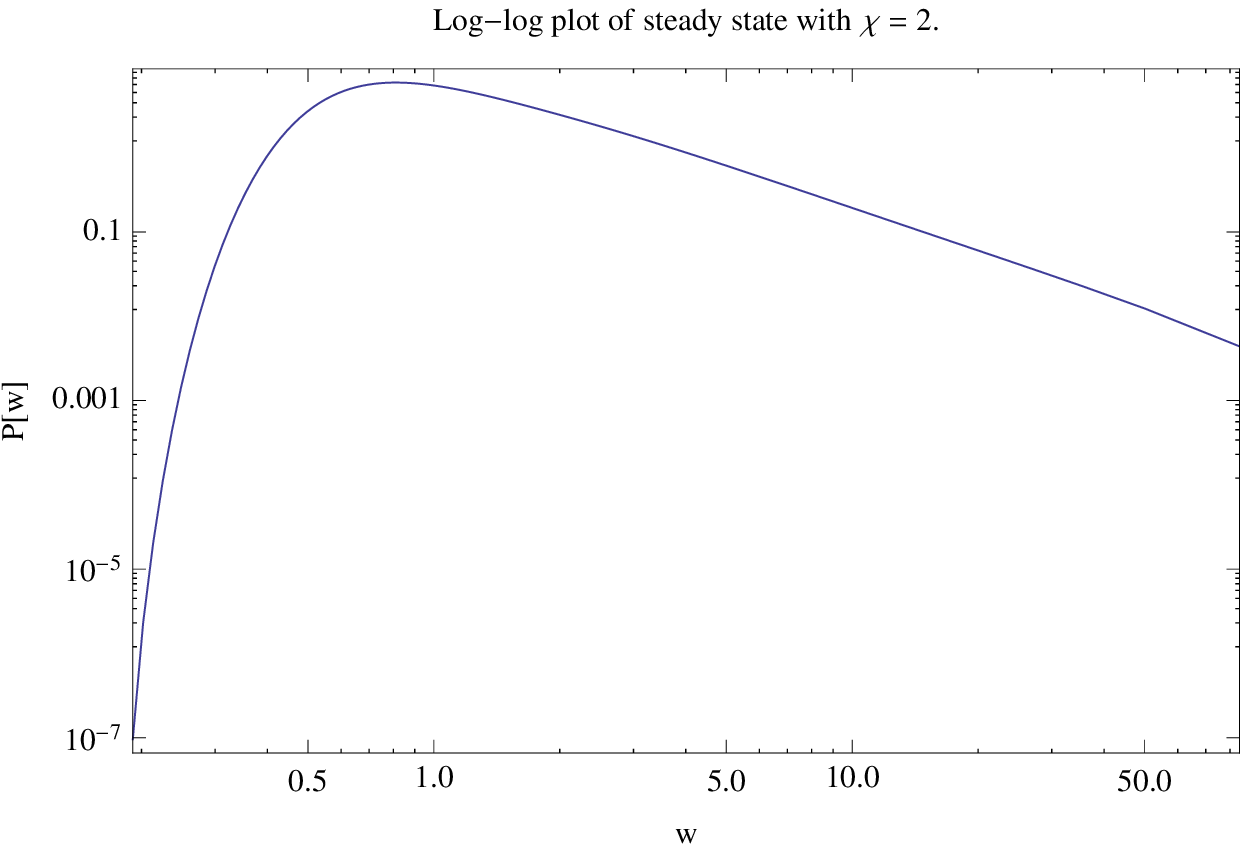}\\
\includegraphics[bbllx=0,bblly=0,bburx=360,bbury=249,width=3.0in]{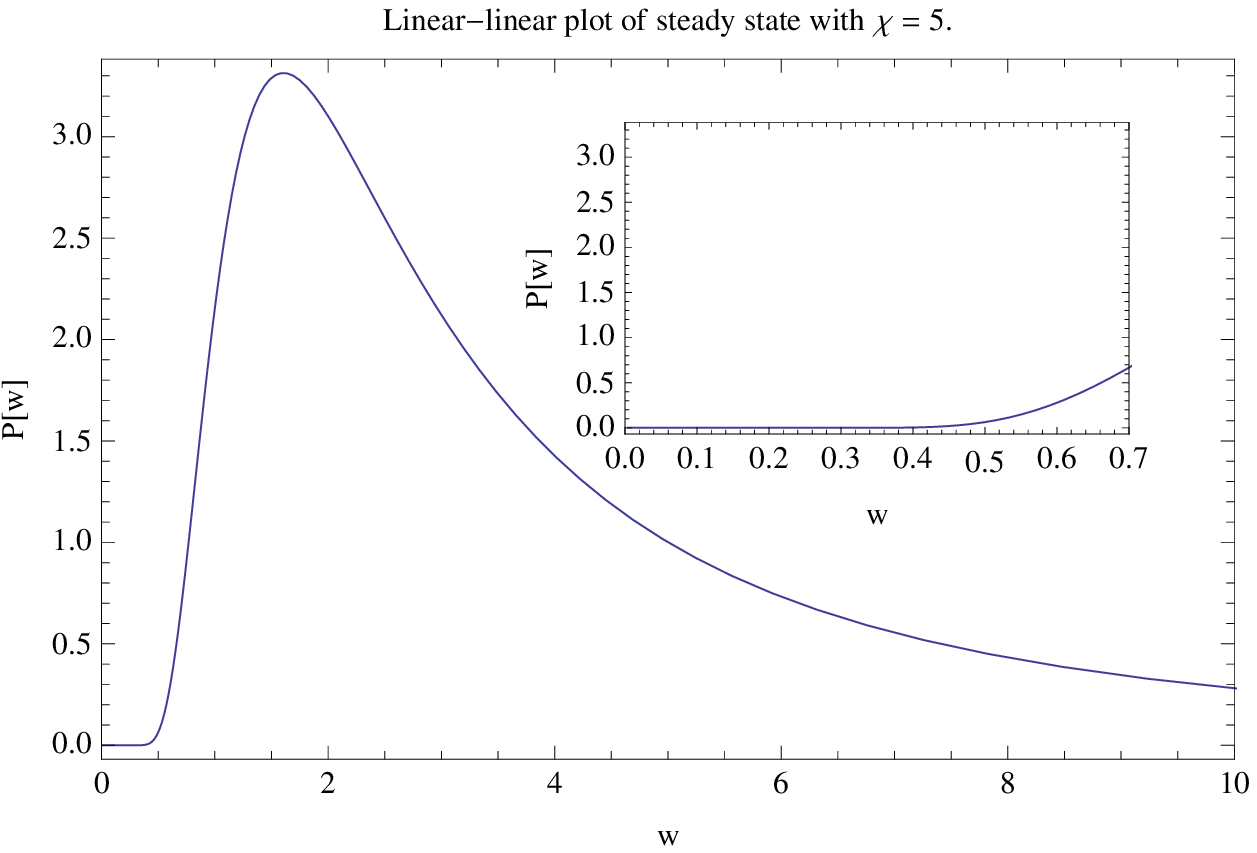}
\hspace{0.1in}
\includegraphics[bbllx=0,bblly=0,bburx=360,bbury=245,width=3.0in]{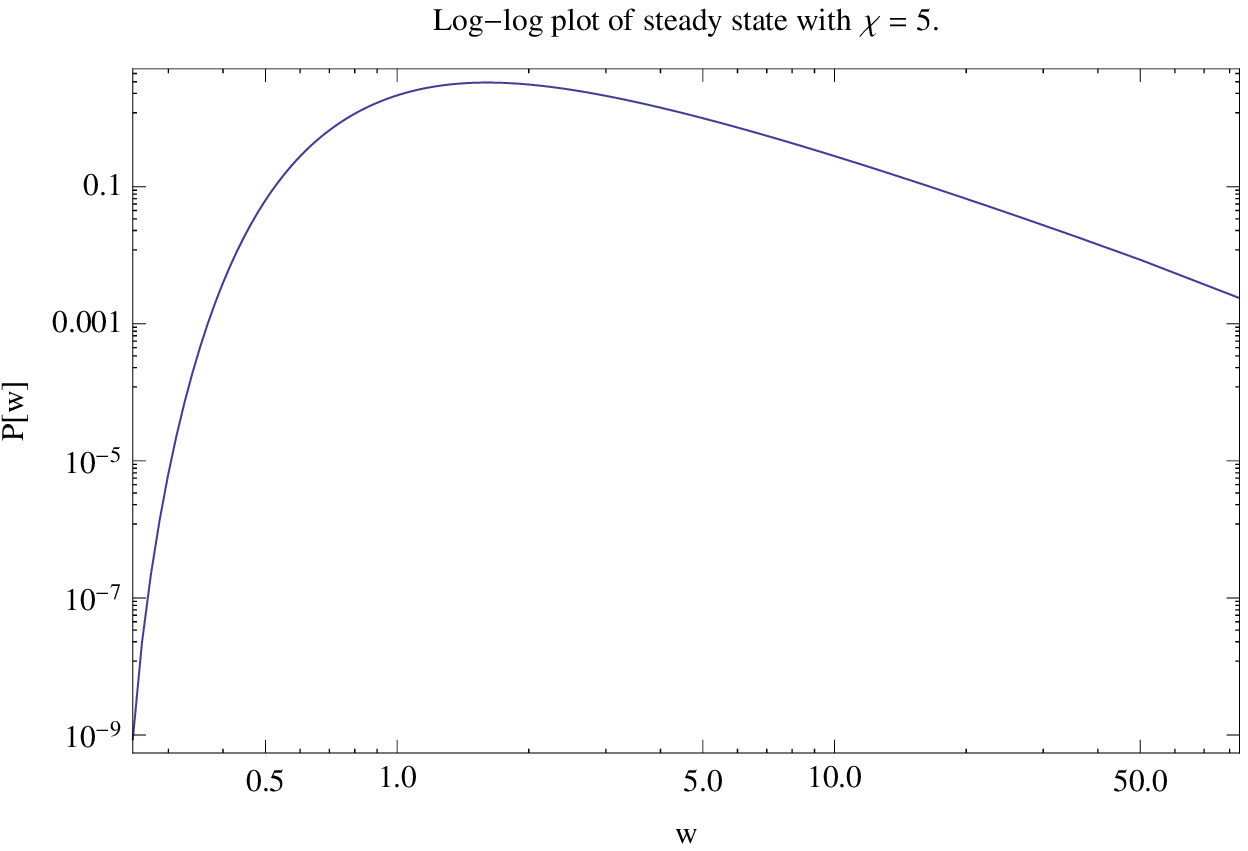}
\end{center}
\caption{{\bf Plots of $P(w)$ versus $w$ for $W/N\approx 7.01126$ and a range of $\chi$:}  This is the steady-state solution computed from Eq.~(\ref{eq:taxationSs}).  The very flat region near the origin in the linear-linear plots is magnified in the insets.  Note that the range of $w$ changes from plot to plot.}
\label{fig:taxes}
\end{figure}

In the steady state, $\partial P/\partial t=0$, Eq.~(\ref{eq:taxationSs}) can be integrated once with respect to $w$ to yield
\begin{equation}
\frac{\partial}{\partial w}\left[
\left(
\frac{w^2}{2}A+B
\right)P
\right] = \chi \left(\frac{W}{N}-w\right) P.
\label{eq:taxationSs}
\end{equation}
Fig.~(\ref{fig:taxes}) shows solutions to the above equation, in both linear-linear and log-log plots, for $W/N=7.01126$ and a range of $\chi$.  The log-log plots make evident behavior at large $w$ that is approximately -- but not exactly -- a power law.  This power-law behavior persists for multiple orders of magnitude.  For smaller values of $\chi$ the curve is noticeably concave up, and for larger values of $\chi$, it is noticeably concave down, indicating deviations from power-law behavior.

Another interesting feature of the plots is the flatness of the solutions near the origin; this feature is emphasized in the insets of the linear-linear plots.  It seems that $P\approx 0$ is very accurate in the vicinity of the origin, and this is very reminiscent of Pareto's cutoff at low $w$, as presented in Eq.~(\ref{eq:ParetoP}).  To see why the solution of Eq.~(\ref{eq:taxationSs}) is very flat near the origin, let us begin by assuming that $P\approx 0$ near the origin, and trying to justify that assumption a posteriori.  It immediately follows that $A\approx 1$ and $B\approx 0$ in this vicinity, so Eq.~(\ref{eq:taxationSs}) reduces to
\begin{equation}
\frac{\partial}{\partial w}\left(
\frac{w^2}{2}P
\right) = \chi \left(\frac{W}{N}-w\right) P.
\label{eq:taxationSsSmallW}
\end{equation}
This has solution
\begin{equation}
P = \frac{C}{w^{2+2\chi}}\exp\left(-2\chi\frac{W}{N}\frac{1}{w}\right),
\label{eq:nonanalytic0}
\end{equation}
where $C$ is a constant of integration.  This function and all its derivatives vanish in the limit as $w\rightarrow 0$, and hence its Taylor series also vanishes, thereby providing us with the desired a posteriori justification to all orders in $w$.  Of course, the function is non-analytic at $w=0$, so this justification completely misses the fact that $P>0$ for $w>0$.  This observation nonetheless suggests that Eq.~(\ref{eq:nonanalytic0}) is approximately valid near the origin, consistent with Pareto's posited cutoff at low $w$.

For large $w$, $A$ and $B$ are no longer well approximated by one and zero, respectively, so we should not expect Eq.~(\ref{eq:nonanalytic0}) to have any validity.  In spite of this, we can choose the integration constant $C$ so that the zeroth and first moments are exactly $N$ and $W$, respectively,
\begin{equation}
P = \frac{N}{\Gamma(2\chi+1)}\left(2\chi\frac{W}{N}\right)^{2\chi+1}
\frac{1}{w^{2+2\chi}}\exp\left(-2\chi\frac{W}{N}\frac{1}{w}\right).
\label{eq:nonanalytic}
\end{equation}
One is tempted to note that the exponential in Eq.~(\ref{eq:nonanalytic}) goes to one for large $w$, leaving us with the power law $w^{-2-2\chi}$, corresponding to Pareto index $\alpha=1+2\chi$.  This is a gross overestimate because $A$ and $B$ are no longer approximately one and zero, respectively, in the power-law regime, but it does capture the fact that the Pareto index $\alpha$ increases with the tax rate $\chi$.  Numerical fits to the data shown in Fig.~\ref{fig:taxes} for $W/N\approx 7.01126$ yield
\begin{center}
\begin{tabular}{cc|cc}
$\chi$ &&& $\alpha$\\
\hline
0.5 &&& 0.49674\\
1.0 &&& 0.53657\\
2.0 &&& 0.64289\\
5.0 &&& 1.03765
\end{tabular}
\end{center}
demonstrating a much more modest increase of $\alpha$ with $\chi$.  Much more work needs to be done to understand the dependence of $\alpha$ on $\chi$ and $W/N$.

\section{Conclusions}
\label{sec:conclude}

The analogy between transacting agents and colliding molecules has been pointed out by a number of authors (see, e.g., Yakovenko~\cite{bib:Yakovenko:2009jt}).  We have pursued this analogy and derived a  general Boltzmann equation governing wealth distribution in the Yard-Sale Model (YSM), with careful attention to all of the assumptions that must go into such a derivation, such as the random-agent approximation.  The construction of this equation so that it conserves both agents and wealth is one of the key results of this paper.

We presented strong analytical and numerical evidence that the dynamics of the YSM make the rich richer and the poor poorer, inexorably driving the distribution of wealth to a decidedly singular state with vanishing Pareto index.  The asymptotic state of the dynamics is one in which all but a vanishingly small fraction of the agents have zero wealth, even while the first moment of the wealth remains positive.  In the Appendix, we introduced the functional analysis necessary to make this last statement rigorous, describing the asymptotic state as a generalized function $\zeta$ which is different from the Dirac delta at zero wealth.

We then introduced the small-transaction limit in which the Boltzmann equation reduces to a simpler partial integrodifferential equation, and presented numerical evidence that this equation has the same singular limit as the Boltzmann equation.  To the best of our knowledge, this PDE has not been posited before in the context of wealth dynamics, and is therefore another of this paper's principal new contributions.

We pointed out that other more detailed artificial society models, such as Sugarscape~\cite{bib:Sugarscape} also exhibit dynamics which drive the Pareto index to values less than unity.  We refuted the usual argument proscribing this, based on the non-normalizability of the wealth distribution.  With lower and upper cutoffs that approach zero and infinity, respectively, at just the correct rates, there is nothing preventing a power-law wealth distribution with Pareto index less than unity.

Finally, we showed how this model may be extended to include phenomena which lead to stationary states with more realistic values of the Pareto index.  In particular, the model with taxation and redistribution of wealth is able to explain something a sharp cutoff at low $w$, due to non-analytic behavior in that vicinity, as well as power-law behavior at large $w$.  This model therefore comes very close to a complete explanation of Pareto's observations.  More detailed analytical and numerical examination of models with these extra features will be the subject of future work~\cite{bib:BUGroup}.

There are many ways in which this work can be expanded and extended.  We can add extra variables to the agents, such as spatial position.  We can examine the development of correlations between transacting agents, and the corrections that these make to the random-agent approximation.  We can examine the possibility of transactions that involve three or more agents at a time, instead of just pairs of agents.  We can also examine steady states of Eq.~(\ref{eq:taxation}).  It is hoped that this presentation will encourage more work along these lines.

\section*{Acknowledgements}

It is a pleasure to thank Harvey Gould, Bill Klein, Kang Liu, Scott MacLachlan, Sid Redner and Jan Tobochnik for helpful discussions during the course of this work.  I would also like to thank David Joulfayan for encouraging me to present this material at the annual meeting of the Armenian Economic Association in Yerevan, Armenia in September 2012.  Most of this work was conducted at the American University of Armenia.  Harvey Gould's careful reading and commentary on an early draft is particularly appreciated, as are the comments of the referees.

\newpage
\bibliography{paper}

\appendix

\section{The Klimontovich representation and inter-agent correlations}
\label{sec:Klimontovich}

In Subsec.~\ref{ssec:adfs}, we introduced the continuous one-agent PDF $P(w)$ and two-agent PDF $P(w,w')$, describing the distribution of wealth amongst a population of agents.  In this appendix we shall make these concepts precise.  In particular, we shall describe the idea, familiar from kinetic theory, that the PDF of wealth $P(w)$ may be understood as the ensemble average of a corresponding quantity in the Klimontovich representation.  As in kinetic theory, we may employ the Klimontovich representation to define multi-agent PDFs, and multi-agent correlation functions.  Because the YSM presented in this paper chooses agents at random, correlations are not a concern; these considerations will become much more important in future versions of this work that account for the manner in which agents are networked.

\subsection{Klimontovich representation of one-agent density function}

We consider a population of $N$ agents with individual wealth $w_j(t)$, where $j=1,\ldots,N$.  The {\it Klimontovich representation} of the one-agent PDF is then
\begin{equation}
P_K(w,t) = \sum_j^N \delta\left(w-w_j(t)\right),
\end{equation}
from which Eqs.~(\ref{eq:N}) and (\ref{eq:W}) yield $N(t) = N$ and $W(t)=\sum_j^N w_j(t),$ respectively.

The Klimontovich representation retains the individual wealth of each agent in the population as a Dirac delta.  For most purposes, this is far too much information to be useful.  The representation that we would prefer is some smoothed version of this.  We may smooth $P_K$ by taking an ensemble average over many different populations of $N$ agents, each evolving independently.  These populations are distinct because their initial conditions may differ and because their time evolution may be stochastic.

To represent the ensemble average mathematically, we add a (possibly multidimensional) ensemble label $\sigma$, so that $w_j(\sigma,t)$ denotes the wealth of the $j$th agent in the $\sigma$th population of the ensemble at time $t$.  For simplicity, we insist that each population in the ensemble has the same number of agents $N$, and the same total wealth $W=\sum_j^N w(\sigma,t)$.  We follow common usage in statistical physics, and refer to an ensemble constructed with these constraints as {\it microcanonical}.  The Klimontovich representation of the one-agent distribution of population $\sigma$ is then denoted
\begin{equation}
P_K(\sigma,w,t) = \sum_j^N \delta\left(w-w_j(\sigma,t)\right).
\end{equation}

The ensemble averaged one-agent distribution is then the integral~\footnote{Note that an average over a finite or countable number of ensemble elements would still yield a singular distribution.  To obtain something smooth, the Dirac deltas of the Klimontovich representation need to be integrated over a continuum.  Some authors avoid this problem by the notational dodge of angle brackets $\langle\cdot\rangle$ for the ensemble average, defined so that $\langle\delta(w-w_j)\rangle$ is somehow smooth.  We eschew this sleight of hand because it evades the real issue:  The Klimontovich distribution is a generalized function, so it belongs inside an integral.  The angle brackets must be the integral over some measure, so it is best to denote them as such.} of this over some measure $d\rho(\sigma)$, normalized so that $\int d\rho(\sigma) = 1$.  That is, the smoothed one-agent PDF that we use is  given by
\begin{equation}
P(w,t) = \int d\rho(\sigma)\; P_K(\sigma,w,t) =
\int d\rho(\sigma)\sum_j^N \delta\left(w-w_j(\sigma,t)\right).
\end{equation}
Because our ensemble is microcanonical, Eqs.~(\ref{eq:N}) and (\ref{eq:W}) still yield $N(t) = N$ and $W(t)=\sum_j^N w_j(\sigma,t),$ respectively, both quantities being independent of $\sigma$.

We note that, in passing from the Klimontovich representation $P_K$ to the smoothed representation $P$, we have lost the discrete nature of $N$ and $W$.  In a real economy, agents are individuals (or other legal entities, such as corporations), and there is necessarily an integer number of them.  Likewise, wealth is measured in some currency, and often rounded off to the minimum unit of that currency, or some rational fraction thereof.  In the smoothed representation, however, $N$ and $W$ are generally real numbers.  We will return to this point later in Subsec.~\ref{ssec:solutions}.

\subsection{Multi-agent density functions}

Similarly, we can define a {\it two-agent density function}~\footnote{It is possible to define a two-agent density function for multiple times as well.  For example, we could define the PDF for finding one agent with wealth $w\in[a,b]$ at time $t$ and another with wealth $w'\in[c,d]$ at time $t'$.  For the purposes of this paper, however, the single-time version with $t'=t$ is all we  need.} at time $t$, denoted by $P(w,w',t)$.  That is, the number of ordered pairs of agents such that one has wealth between $a$ and $b$ and the other has wealth between $c$ and $d$ at time $t$ is given by $\int_a^b dw\int_c^d dw'\; P(w,w',t)$.  This two-agent PDF satisfies three important properties:
\begin{enumerate}
\item[(i)] Because the total number of ordered pairs of agents is $N^2$, we must have~\footnote{Note that, because these are {\it ordered} pairs, we count the pairing of an agent with wealth $w$ with another with wealth $w'$ as distinct from the reverse.  We also include pairings of agents with themselves.  This is why the total number of pairs is $N^2$.}
\begin{equation}
N^2 = \int_0^\infty dw\int_0^\infty dw'\; P(w,w',t).
\end{equation}
\item[(ii)] Because the property of being paired is symmetric, we must have
\begin{equation}
P(w,w',t) = P(w',w,t).
\end{equation}
\item[(iii)] Because each agent may be paired with $N$ others, integrating the two-agent PDF over the second variable and dividing by $N$ must yield the one-agent density function,
\begin{equation}
P(w,t) = \frac{1}{N}\int_0^\infty dw'\; P(w,w',t).
\label{eq:projection}
\end{equation}
\end{enumerate}

To better understand the two-agent PDF, we first consider its Klimontovich representation
\begin{equation}
P_K(w,w',t) = \sum_j^N\sum_k^N \delta\left(w-w_j(t)\right)\delta\left(w'-w_k(t)\right).
\end{equation}
It is manifest that this factors,
\begin{equation}
P_K(w,w',t) = P_K(w,t)P_K(w',t),
\end{equation}
so that the Klimontovich representation of the two-agent PDF is the product of two one-agent Klimontovich PDFs.  With this observation, the three properties in the foregoing paragraph follow immediately.

As with the one-agent PDF, the Klimontovich representation of the two-agent PDF contains much more information than we need, so we smooth it by taking an ensemble average,
\begin{equation}
P(w,w',t) = \int d\rho(\sigma)\; P_K(\sigma,w,w',t) =
\int d\rho(\sigma)\sum_j^N\sum_k^N
\delta\left(w-w_j(\sigma,t)\right)\delta\left(w'-w_k(\sigma,t)\right).
\end{equation}
As a consequence of the ensemble average, the smoothed two-agent PDF no longer factors into a product form, but we can write
\begin{equation}
P(w,w',t) = P(w,t)P(w',t) + C(w,w',t),
\label{eq:correlation}
\end{equation}
where we have defined the two-agent correlation function
\begin{equation}
C(w,w',t) := \int d\rho(\sigma)
\left(\sum_j^N\delta\left(w-w_j(\sigma,t)\right) - P(w,t)\right)
\left(\sum_k^N\delta\left(w'-w_k(\sigma,t)\right) - P(w',t)\right),
\end{equation}
which may be thought of as the excess probability of finding a pair of agents, over and above the product of the probabilities of finding each individually.  As with one-agent PDFs, we sometimes suppress the time dependence, writing for example $P(w,w')$ and $C(w,w')$, instead of $P(w,w',t)$ and $C(w,w',t)$, if the time is obvious from the context.

It follows from the definition of the two-agent correlation function that
\begin{equation}
0 = \int_0^\infty dw\; C(w,w',t) = \int_0^\infty dw'\; C(w,w',t)
\end{equation}
and
\begin{equation}
C(w,w',t) = C(w',w,t),
\end{equation}
and from these one can verify that $P(w,w',t)$ still satisfies properties (i) through (iii) above, even though it is no longer a product form.

Likewise, $p$-agent PDFs for $p>2$ can also be expressed as product forms supplemented by connected correlation functions.

\section{Conservation laws for the Boltzmann equation}
\label{sec:conservationLawsBoltzmann}

In this Appendix, we demonstrate that the quantities $N$ and $W$, defined in Eqs.~(\ref{eq:N}) and (\ref{eq:W}), are constants of the motion of Eq.~(\ref{eq:Boltzmann}).

\subsection{Conservation of agents}

To demonstrate that the total number of agents, given by Eq.~(\ref{eq:N}), is conserved, we first note that
\begin{eqnarray}
\frac{dN}{dt}
&=&
\frac{d}{dt}
\int_0^\infty dw\; P(w,t)
=
\int_0^\infty dw\; \frac{\partial P(w,t)}{\partial t}
\nonumber\\
&=&
\int_{-1}^{+1} d\beta\; \eta(\beta)
\left\{
-\int_0^\infty dw\left[
P(w,t)
-
\frac{1}{1+\beta}
P\left(\frac{w}{1+\beta},t\right)
\right]\right.
\nonumber\\
& &
+
\frac{1}{N}
\int_0^\infty dw
\int_0^{\frac{w}{1+\beta}} dw'\;
\left.
\left[
P\left(w - \beta w',t\right)
-
\frac{1}{1+\beta}
P\left(\frac{w}{1+\beta},t\right)
\right]P(w',t)
\right\},
\label{eq:consNa}
\end{eqnarray}
where we have exchanged the order of integration over $\beta$ and $w$ in the second line.  It follows that $N$ will be conserved if the right-hand side vanishes.  In fact, we will show that the two terms in the curly brackets vanish separately.

First, we note that a simple change of integration variable in the first term establishes that
\begin{equation}
\int_0^\infty dw\;
\left[
P(w,t) -
\frac{1}{1+\beta}
P\left(\frac{w}{1+\beta},t\right)
\right]
=
0.
\label{eq:consNb}
\end{equation}
Next, we note that
\begin{eqnarray}
\lefteqn{
\frac{1}{N}
\int_0^\infty dw\;
\int_0^{\frac{w}{1+\beta}} dw'\;
\left[
P\left(w - \beta w',t\right)
-
\frac{1}{1+\beta}
P\left(\frac{w}{1+\beta},t\right)
\right]
P\left(w',t\right)}
\nonumber\\
&=&
\frac{1}{N}
\int_0^\infty dw'\;
P\left(w',t\right)
\int_{(1+\beta)w'}^\infty dw\;
\left[
P\left(w - \beta w',t\right)
-
\frac{1}{1+\beta}
P\left(\frac{w}{1+\beta},t\right)
\right]
\nonumber\\
&=&
\frac{1}{N}
\int_0^\infty dw'\;
P\left(w',t\right)
\left[
\int_{w'}^\infty dw\;
P\left(w,t\right)
-
\int_{w'}^\infty dw\;
P\left(w,t\right)
\right]
\nonumber\\
&=&
0,
\label{eq:consNc}
\end{eqnarray}
where we have changed the order of integration in the first step, and made two different substitutions in the second step.  Combining Eqs.~(\ref{eq:consNa}), (\ref{eq:consNb}) and (\ref{eq:consNc}), we find
\begin{equation}
\frac{dN}{dt} = 0,
\end{equation}
as expected.

\subsection{Conservation of wealth}

Likewise, to demonstrate that the total wealth of the population, given by Eq.~(\ref{eq:W}), is conserved, we first note that
\begin{eqnarray}
\frac{dW}{dt}
&=&
\frac{d}{dt}
\int_0^\infty dw\; w P(w,t)
=
\int_0^\infty dw\; w \frac{\partial P(w,t)}{\partial t}
\nonumber\\
&=&
\int_{-1}^{+1} d\beta\; \eta(\beta)
\left\{
-\int_0^\infty dw\; w\left[
P(w,t)
-
\frac{1}{1+\beta}
P\left(\frac{w}{1+\beta},t\right)
\right]\right.
\nonumber\\
& &
+
\frac{1}{N}
\int_0^\infty dw\; w
\int_0^{\frac{w}{1+\beta}} dw'\;
\left.
\left[
P\left(w - \beta w',t\right)
-
\frac{1}{1+\beta}
P\left(\frac{w}{1+\beta},t\right)
\right]P(w',t)
\right\},
\label{eq:consWa}
\end{eqnarray}
where we have exchanged the order of integration over $\beta$ and $w$ in the second line.  It follows that $W$ will be conserved if the right-hand side vanishes.  This time, we shall show that the two terms in curly brackets are both odd functions of $\beta$, so that when they are integrated along with the even function $\eta(\beta)$, the result vanishes.

A simple change of integration variable in the first term establishes that
\begin{equation}
\int_0^\infty dw\; w
\left[
P(w,t)
-
\frac{1}{1+\beta}
P\left(\frac{w}{1+\beta},t\right)
\right]
=
-\beta W,
\label{eq:consWb}
\end{equation}
which is proportional to $\beta$.  We also have that
\begin{eqnarray}
\lefteqn{
\frac{1}{N}
\int_0^\infty dw\; w
\int_0^{\frac{w}{1+\beta}} dw'\;
\left[
P\left(w - \beta w',t\right)
-
\frac{1}{1+\beta}
P\left(\frac{w}{1+\beta},t\right)
\right]
P\left(w',t\right)}
\nonumber\\
&=&
\frac{1}{N}
\int_0^\infty dw'\;
P\left(w',t\right)
\int_{(1+\beta)w'}^\infty dw\; w
\left[
P\left(w - \beta w',t\right)
-
\frac{1}{1+\beta}
P\left(\frac{w}{1+\beta},t\right)
\right]
\nonumber\\
&=&
\frac{1}{N}
\int_0^\infty dw'\;
P\left(w',t\right)
\left[
\int_{w'}^\infty dw\; (w+\beta w')
P\left(w,t\right)
-
(1+\beta)
\int_{w'}^\infty dw\; w
P\left(w,t\right)
\right]
\nonumber\\
&=&
\frac{\beta}{N}
\int_0^\infty dw'\;
P\left(w',t\right)
\int_{w'}^\infty dw\;
P\left(w,t\right) (w' - w),
\label{eq:consWc}
\end{eqnarray}
which is also proportional to $\beta$.  In Eq.~(\ref{eq:consWc}), we have changed the order of integration in the first step, and made two different substitutions in the second step.  Combining Eqs.~(\ref{eq:consWa}), (\ref{eq:consWb}) and (\ref{eq:consWc}), and invoking the evenness of the function $\eta(\beta)$, we find
\begin{equation}
\frac{dW}{dt} = 0,
\end{equation}
as expected.

Note that wealth conservation follows from the average of the rates of change for the winning and losing scenarios, reflected in the evenness of $\eta(\beta)$, as described in the discussion leading from Eq.~(\ref{eq:halfBBGKY}) to Eq.~(\ref{eq:halfBBGKY}).  Wealth is not conserved by the winning and losing scenarios separately.  For example, the foregoing argument should make it clear that Eq.~(\ref{eq:halfBBGKY}), by itself, does not conserve total wealth.

\section{Description of the time-asymptotic limit}
\label{sec:GenFun}

As noted in the text, the function $P(w,t)$ and its approximation $P_c(w,t)$ approach a generalized function as $t\rightarrow\infty$.  This generalized function has support only at the origin, and has zeroth moment equal to $N$.  This suggests the limit $N\delta(w)$, but we additionally require that it have first moment $W$.  In the function space $L_2$, this additional requirement is impossible to satisfy.  We are forced to the conclusion that the dynamics of wealth can evolve $P(w,t)$ to something outside $L_2$ in the $t\rightarrow\infty$ limit.  The appropriate function space in which to study the time asymptotics of wealth is therefore a larger function space than $L_2$.  This Appendix describes the functional analysis that is necessary to make this statement rigorous.  The discussion is meant to be self-contained, requiring little prior background in the subject.

Our numerical simulations clearly indicate that the asymptotic state of the system has $N-1$ agents in a state of abject poverty, and one with all the wealth $W$.  As noted in the text, however, this division between $N-1$ poor agents and one wealthy agent is due to the discrete nature of the simulation.  If we could simulate the continuous distribution of agents governed by Eq.~(\ref{eq:Boltzmann}), we might expect to see ever smaller ``fractions of agents'' $f$ with wealth $W/f$, alongside $N-f$ agents living in poverty.  If $f\rightarrow 0$ in the time-asymptotic limit, we might expect that everybody eventually ends up poor, in some sense, so that a good generalized function candidate for $\lim_{t\rightarrow\infty}P(w,t)$ or $\lim_{t\rightarrow\infty}P_c(w,t)$ might be $N\delta(w)$.  Indeed, this view is reinforced by noting that Eq.~(\ref{eq:Boltzmann}) can be rewritten in the suggestive form
\begin{eqnarray}
\lefteqn{
\frac{\partial P(w,t)}{\partial t}
}\nonumber\\
&=&
\int_{-1}^{+1} d\beta\; \eta(\beta)
\frac{1}{N}
\int_{-0}^{\frac{w}{1+\beta}} dw'\;
\left[
P\left(w - \beta w',t\right)
-
\frac{1}{1+\beta}
P\left(\frac{w}{1+\beta},t\right)
\right]
\left[
P(w',t) -
N\delta(w')
\right],\nonumber\\
\end{eqnarray}
where the notation $-0$ for the lower limit of integration is meant to emphasize that the Dirac delta is entirely contained within the region of integration.  This form makes clear that $P(w,t)=N\delta(w)$ is a steady state solution of Eq.~(\ref{eq:Boltzmann}).  It is also zero for $w>0$, consistent with the $a\rightarrow 0$ limit of $a/w$.  Unfortunately, though it obviously satisfies Eq.~(\ref{eq:N})~\footnote{with a lower limit of integration of $-0$ as above}, it does not satisfy Eq.~(\ref{eq:W}), except in the trivial economy that has $W=0$.  This will not do.

One's next guess might be $P(w,t)=N\delta(w) - W\delta'(w)$, since this satisfies both Eq.~(\ref{eq:N}) and Eq.~(\ref{eq:W}).  Here the trouble is that all higher moments of this generalized function, such as $M=\int w^2 P(w)\; dw$, vanish.  In reality, these higher moments are either nonzero or divergent.

We need a generalized function $\zeta(w)$, defined for $w\in[0,\infty)$, that has the following four properties:
\begin{enumerate}
\item[(i)] $\zeta(w) = 0$ for $w> 0$
\item[(ii)] $\int_0^\infty dw\;\zeta(w) = N$
\item[(iii)] $\int_0^\infty dw\;\zeta(w) w = W$
\item[(iv)] $\int_0^\infty dw\;\zeta(w) w^j>0$ for $j\geq 2$.
\end{enumerate}
The first two of these are reminiscent of the ``physicists' definition'' of ($N$ times) a Dirac delta.  As is well known, the apparent absurdity of these simultaneous demands was resolved mathematically only by the advent of the theory of distributions by Sobolev, Schwartz and others between the 1930s and the 1950s~\cite{bib:Griffel}.  The question facing us now is how to use distribution theory to define a generalized function $\zeta$ with {\it all four} of the above properties.

Distribution theory requires a space $\calD$ of {\it test functions} $\psi(w)$ that are smooth and have bounded support~\footnote{A function is {\it smooth} if it is infinitely differentiable.  A function has {\it bounded support} if the set of $w$ for which it is nonzero (more precisely, the closure of that set) is a subset of $[a,b]$ for some real  $a$ and $b$.}.  Generalized functions are then associated with linear functionals on this space.  The action of a functional $f$ on a test function $\psi$ is a map $\calD\rightarrow {\mathbb R}$, and the real number that results is usually denoted $\langle f,\psi\rangle$.  For example, the functional $\delta$ defined by $\langle\delta,\psi\rangle=\psi(0)$ is the Dirac delta.  It is easily seen to be a linear functional, since
\begin{equation}
\langle\delta,c_1\psi_1+c_2\psi_2\rangle =
(c_1\psi_1+c_2\psi_2)(0) =
c_1\psi_1(0)+c_2\psi_2(0) =
c_1\langle\delta,\psi_1\rangle + c_2\langle\delta,\psi_2\rangle.
\end{equation}
In this way of thinking, $\delta$ is not a function of $w$; rather, it is a functional on $\calD$.  We may then revert to writing $\int_0^\infty dw\;\delta(w)\psi(w)$ in place of $\langle\delta,\psi\rangle$, but it must be understood that this is an abuse of notation.  There is never any question about what the value of $\delta(w)$ is at a particular $w$.  Whenever ambiguity arises, we turn to the interpretation of $\delta$ as a linear functional on $\calD$ to resolve it.  An excellent introduction to distribution theory may be found in, for example, the first few chapters of the text by Griffel~\cite{bib:Griffel}.

To put the generalized function $\zeta$ on a firm footing, we need more requirements on our space of test functions.  Let us first consider the space $\calG$ of test functions $\psi$ that are smooth and have bounded support on $[0,\infty)$, and for which
\begin{equation}
F[\psi] := \int_0^\infty dw\;\frac{\left|\psi(w)-\psi(0)\right|}{w} < \infty.
\label{eq:Gcondition}
\end{equation}
The reader may verify, for example, that the test function
\begin{equation}
\psi(w) = \left\{
\begin{array}{ll}
\exp\left[-\frac{1}{w(1-w)}\right] & \mbox{for $0<w<1$}\\
0 & \mbox{otherwise}
\end{array}
\right.
\end{equation}
belongs to $\calG$.  By contrast, the functions $\psi(w)=1$ and $\psi(w)=w$ do not belong to $\calG$, because they do not have bounded support; in the latter case, there is also the problem that $F$ applied to $\psi$ is not finite.

We should first verify that $\calG$ is indeed a linear space.  We do this by supposing that we have two test functions $\psi_1\in\calG$ and $\psi_2\in\calG$.  This means that both $\psi_1$ and $\psi_2$ are smooth and have bounded support on $[0,\infty)$, and that $F[\psi_j] < \infty$ for $j=1,2$.  We now consider the linear combination $c_1\psi_1+c_2\psi_2$.  It is clear that this combination is also smooth and has bounded support on $[0,\infty)$.  We then note that the linear combination satisfies Eq.~(\ref{eq:Gcondition}), since
\begin{eqnarray}
F[c_1\psi_1+c_2\psi_2]
&=&
\int_0^\infty dw\;\frac{\left|c_1\psi_1(w)+c_2\psi_2(w)-c_1\psi_1(0)-c_2\psi_2(0)\right|}{w}
\nonumber\\
&\leq&
\int_0^\infty dw\;\frac{\left|c_1\psi_1(w)-c_1\psi_1(0)\right|+\left|c_2\psi_2(w)-c_2\psi_2(0)\right|}{w}
\nonumber\\
&\leq&
|c_1|F[\psi_1]+
|c_2|F[\psi_2]
\nonumber\\
&<&
\infty,
\label{eq:linearity}
\end{eqnarray}
where we have used the triangle inequality.  So $\calG$ is closed under linear combinations, and thereby qualifies as a linear space.

In fact, $\calG$ is not quite big enough for our purposes.  We want the functions $\phi(w)=1$ and $\phi(w)=w$ and constant multiples thereof to be in our space of test functions, but, as noted above, they are not in $\calG$.  So we next define $\calG_1$ to be the space of functions that are the sum of a function in $\calG$ and any linear function of $w$.  That is, for each $\phi\in\calG_1$, we may write $\phi(w) = \psi(w) + \gamma + \mu w$, where $\psi\in\calG$ and $\gamma,\mu\in{\mathbb R}$.  Moreover, we shall demonstrate that this decomposition is unique.  For any function $\phi\in\calG_1$ there are unique real numbers $\gamma$ and $\mu$, such that $\psi(w) = \phi(w) - \gamma - \mu w\in\calG$.

Before showing how to compute $\gamma$ and $\mu$, we should make an incidental comment:  The principal reason for using test functions with bounded support in distribution theory is to allow us to integrate by parts, discarding surface terms with reckless abandon.  Note that we can do this in the space $\calG$, but we will need to be a bit more careful in the space $\calG_1$ because $\lim_{w\rightarrow\infty}\phi'(w) = \mu$.

To calculate $\gamma$ and $\mu$ from $\phi\in\calG_1$, note that $\lim_{w\rightarrow\infty}\left(\phi(w)-\mu w\right)=\gamma$ follows from the fact that $\psi$ has bounded support.  The approach is then to show that $\mu$ is the unique real number for which the limit $\lim_{w\rightarrow\infty}\left(\phi(w)-\mu w\right)$ exists, and that for this value of $\mu$ the value of the limit is $\gamma$.

To see that this approach defines $\mu$ uniquely, let us suppose that there were two values $\mu_1$ and $\mu_2$ for which the limit existed.  That is, suppose that
\begin{eqnarray}
\lim_{w\rightarrow\infty}\left(\phi(w)-\mu_1 w\right) &=& \gamma_1\\
\lim_{w\rightarrow\infty}\left(\phi(w)-\mu_2 w\right) &=& \gamma_2
\end{eqnarray}
are both finite and real.  Since both limits exist, we can subtract these equations to obtain
\begin{equation}
\lim_{w\rightarrow\infty}\left[\left(\mu_2-\mu_1\right) w\right] = \gamma_1-\gamma_2,
\end{equation}
but there is no way that this last statement can be true, unless $\mu_1=\mu_2$.  Uniqueness of $\gamma$ then follows immediately.

The unique association of $\phi\in\calG_1$ with the constant $\mu$, such that $\lim_{w\rightarrow\infty}\left(\phi(w)-\mu w\right)$ exists, is itself a linear functional, which we shall call $\Xi$; that is, we write $\langle\Xi,\phi\rangle = \mu$.  To demonstrate linearity of $\Xi$, let us suppose that $\phi_j\in\calG_1$, so that $\lim_{w\rightarrow\infty}\left(\phi_j(w)-\mu_j w\right)=\gamma_j$ exists, and we can write $\langle\Xi,\phi_j\rangle = \mu_j$ for $j=1,2$.  By taking a linear combination of these limits, it follows that
\begin{equation}
\lim_{w\rightarrow\infty}
\left[\left(c_1\phi_1(w)+c_2\phi_2(w)\right)-\left(c_1\mu_1+c_2\mu_2\right) w\right]=c_1\gamma_1+c_2\gamma_2
\end{equation}
also exists, so
\begin{equation}
\langle\Xi,c_1\phi_1+c_2\phi_2\rangle = c_1\mu_1+c_2\mu_2=
c_1\langle\Xi,\phi_1\rangle+c_2\langle\Xi,\phi_2\rangle,
\label{eq:linearity2}
\end{equation}
thereby demonstrating linearity of the functional $\Xi$ and justifying our notation.

Armed with our space ${\calG_1}$ of test functions and the functional $\Xi$, we are now ready to make sense of the generalized function $\zeta$, described earlier.  In the language of distributions, $\zeta$ may be written
\begin{equation}
\zeta = N\delta + W\Xi.
\end{equation}
That is, for any test function $\phi\in\calG_1$, where $\phi(w) = \psi(w)+\gamma +\mu w$ with $\psi\in\calG$, we have
\begin{equation}
\langle\zeta,\phi\rangle = N\phi(0) + W\mu.
\end{equation}
As with $\delta$, we may now abuse notation by writing the above as follows
\begin{equation}
\int_0^\infty dw\;\zeta(w)\phi(w) = N\phi(0) + W\mu.
\end{equation}
Setting $\phi(w)=1$, we find $\gamma=1$ and $\mu=0$, so it follows that
\begin{equation}
\int_0^\infty dw\;\zeta(w) = N.
\end{equation}
Setting $\phi(w)=w$, we find $\gamma=0$ and $\mu=1$, so it follows that
\begin{equation}
\int_0^\infty dw\;\zeta(w) w = W.
\end{equation}
Thus, the generalized function $\zeta$ satisfies Eqs.~(\ref{eq:N}) and (\ref{eq:W}).

Note that ${\calG_1}$ may be characterized as the space of functions whose second derivative is in ${\calG}$.  (If we have $\phi(w) = \psi(w) + \gamma + \mu w$, then clearly $\phi(w)$ and $\psi(w)$ have the same second derivative.)  This observation relates ${\calG_1}$ to a class of function spaces known as {\it Sobolev spaces}, but elaboration of this point would take us beyond the scope of this paper.

Can we now prove that the function $P_c$, defined in Eq.~(\ref{eq:Pc}), {\it converges weakly} to $\zeta$ in the limit as $a\rightarrow 0$?  For an arbitrary $\phi\in\calG_1$, and for $\mu=\langle\Xi,\phi\rangle$ and $\psi(w)=\phi(w)-\gamma-\mu w\in\calG$, we consider the quantity
\begin{eqnarray}
\left|\langle P_c - \zeta,\phi\rangle\right|
&=&
\left|
\int_0^\infty dw\; \left[P_c(w)-\zeta(w)\right]\phi(w)
\right|
\nonumber\\
&=&
\left|
\int_0^\infty dw\; P_c(w)\left[\psi(w)+\gamma+\mu w\right] - N\phi(0) - W\mu
\right|
\nonumber\\
&=&
\left|
\int_0^\infty dw\; P_c(w)\left[\psi(w)-\psi(0)\right]
\right|
\nonumber\\
&=&
\left|
a\int_{a_{\min}}^{a_{\max}} dw\; \frac{\psi(w)-\psi(0)}{w}
\right|
\nonumber\\
&\leq&
|a|
\int_{a_{\min}}^{a_{\max}} dw\; \frac{\left|\psi(w)-\psi(0)\right|}{w}
\nonumber\\
&\leq&
|a|
\int_0^\infty dw\; \frac{\left|\psi(w)-\psi(0)\right|}{w}
\nonumber\\
&\leq&
M|a|,
\end{eqnarray}
where $M=F[\psi]<\infty$ because $\psi\in\calG$, and where we used the fact that $P_c$ was constructed to obey Eqs.~(\ref{eq:N}) and (\ref{eq:W}).  It follows that
\begin{equation}
\lim_{a\rightarrow 0}\left|\langle P_c - \zeta,\phi\rangle\right| = 0.
\label{eq:this6}
\end{equation}
Because Eq.~\eqref{eq:this6} holds for arbitrary test functions $\phi\in\calG_1$, we can conclude that $P_c$ converges weakly to $\zeta$ in the limit as $a\rightarrow 0$ or $t\rightarrow\infty$ in the function space $\calG_1$.  Our numerical evidence then strongly suggests that $P$ obeying Eq.~(\ref{eq:Boltzmann}) likewise converges weakly to $\zeta$.  This last point is, of course, not proven by the above arguments, but we offer it as a very plausible conjecture.

Finally, we can show that the generalized function $\zeta(w)$ described above is also a weak stationary state of the dynamical equation for the small-transaction limit, Eq.~(\ref{eq:smallBeta}).  To see this, we examine the integral of the right-hand side of Eq.~(\ref{eq:smallBeta}) multiplied by an arbitrary test function $\phi\in\calG_1$,
\begin{equation}
\int_0^\infty
\phi(w)\frac{\partial^2}{\partial w^2}
\left[
\left(\frac{w^2}{2}A+B\right)\zeta
\right]\; dw.
\end{equation}
Writing $\phi(w) = \psi(w) + \gamma + \mu w$ as before, and integrating by parts twice, we find
\begin{equation}
\int_0^\infty
\frac{\partial^2\psi(w)}{\partial w^2}
\left(\frac{w^2}{2}A+B\right)\zeta
\; dw.
\end{equation}
To evaluate this last integral, let $f(w):= \psi''(w)(w^2 A/2 + B)$.  Then the integral is equal to $Nf(0)+W\mu$, where $\mu$ is the unique number such that $\lim_{w\rightarrow\infty}\left[f(w)-\mu w\right]$ exists.  We first note that $f(0)=0$ because $w^2 A/2+B$ vanishes at $w=0$ (and $\psi$ is smooth).  Then $\mu=0$ follows from the fact that $\psi$ has bounded support.  So the integral vanishes, and the generalized function $\zeta(w)$ is a stationary state of Eq.~(\ref{eq:smallBeta}) in this weak sense.  We conjecture that it is the stationary state to which arbitrary initial conditions generically attract.

\end{document}